\documentclass[pre,aps,twocolumn,showpacs,floatfix,10pt,reprint]{revtex4-1}
\usepackage[dvips]{graphicx}
\usepackage{amsmath}
\usepackage{amssymb}
\usepackage[nice]{nicefrac}
\usepackage{color}
\usepackage[hidelinks]{hyperref}
\hypersetup{colorlinks=false}
\newcommand{\bd}[1]{\textcolor{black}{#1}}

\begin{document}

\title{L\'evy flights versus L\'evy walks in bounded domains}

\author{Bart{\l}omiej Dybiec}
\email{bartek@th.if.uj.edu.pl}
\affiliation{Marian Smoluchowski Institute of Physics, and Mark Kac Center for Complex Systems Research, Jagiellonian University, ul. St. {\L}ojasiewicza 11, 30--348 Krak\'ow, Poland}

\author{Eli Barkai}
\email{eli.barkai@biu.ac.il}
\affiliation{Department of Physics, Bar Ilan University, Ramat-Gan 52900, Israel}

\author{Alexander A. Dubkov}
\email{dubkov@rf.unn.ru}
\affiliation{Radiophysical Department, Lobachevsky State University, Gagarin ave. 23, 603950 Nizhni Novgorod, Russia}

\author{Ewa Gudowska-Nowak}
\email{gudowska@th.if.uj.edu.pl}
\affiliation{Marian Smoluchowski Institute of Physics, and Mark Kac Center for Complex Systems Research, Jagiellonian University, ul. St. {\L}ojasiewicza 11, 30--348 Krak\'ow, Poland}

\date{March 8, 2017}
\begin{abstract}
L\'evy flights and L\'evy walks serve as two paradigms of random walks resembling common features but also bearing fundamental differences. One of the main dissimilarities are discontinuity versus continuity of their trajectories and infinite versus finite propagation velocity. In consequence, well developed theory of L\'evy flights is associated with their pathological physical properties, which in turn are resolved by the concept of L\'evy walks. Here, we explore L\'evy flights and L\'evy walks models on bounded domains examining their differences and analogies.
We investigate analytically and numerically whether and under which conditions both approaches yield similar results in terms of selected statistical observables characterizing the motion: the survival probability, mean first passage time and stationary PDFs.
It is demonstrated that similarity of models is affected by the type of boundary conditions and value of the stability index defining asymptotics of the jump length distribution.

\end{abstract}

\maketitle



\section{Introduction}

L\'evy flights \cite{shlesinger1995} and L\'evy walks \cite{klafter1996,zaburdaev2015levy} are two well known stochastic models of anomalous diffusion.
Generally speaking, L\'evy flights correspond to Markovian motions whose individual, uncorrelated random steps are drawn from a L\'evy distribution, thus extending Brownian motion for which the step lengths are Gaussian. A resulting asymptotic L\'evy diffusion is then characterized by infinite variance, indicating that the width of the diffusive ``packet'' must be understood in terms of some fractional moments or the interquartile distance \cite{dybiec2008d}. This mathematical property of L\'evy flights along with their instantaneous propagation
are considered in many situations unphysical.
In contrast, in L\'evy walks \cite{shlesinger1986,klafter1996,zaburdaev2015levy}, the meandering particle has a finite
velocity, so that long jumps take proportionally longer time. Still, in the absence of any boundary effect, the core of the L\'evy walk packet
\bd{disperses faster than linearly in time but slower than the ballistic front and is described by the L\'evy distribution}.
This means that under free boundary conditions L\'evy flights can serve as a good approximation to the L\'evy walk, although with an improper prediction of the moments of the jump length distribution \cite{zaburdaev2015levy}.

In this paper we analyze these two popular models of stochastic motion in bounded domains.
Such motions can e.g. represent foraging behaviors of animals and bacteria \cite{zaburdaev2015levy,metzler2012levy}, the spreading of diseases \cite{hufnagel2004} or particle transport along soft polymer chains \cite{sokolov1997}.
The problem we aim to address
is whether both aforementioned approaches yield similar results in terms of investigated kinetics (survival, occupation and first passage times) and
long-term behavior (existence of stationary states and stationary probability densities).
Depending on the value of the stability index $\alpha$ and used characteristics we find similarities but also large deviations between the two models.
We furthermore compare explicit analytical results with numerical simulations of stochastic dynamics.
Direct comparison of two models -- motion of random walkers flying instantaneously between distinct sites and walkers performing motion at a constant speed -- deepens our understanding of their behavior and relates
characteristic properties amenable to measurements in real situations.

The article is organized as follows: Sec.~\ref{sec:lf} discuses the problem of boundary conditions, mean first passage time, mean residence time and stationary states for L\'evy flights.
In Sec.~\ref{sec:lw} problems of boundary conditions, mean first passage time and stationary states for L\'evy walks are explored.
The manuscript is closed with summary and discussion.

\section{L\'evy flights in one dimensional intervals\label{sec:lf}}

Let us briefly reconsider the motion of a free overdamped particle described by the Langevin equation:
\begin{equation}
 \frac{dx}{dt}=\zeta_\alpha(t),
 \label{eq:langevin}
\end{equation}
where $\zeta_\alpha(t)$ is a symmetric white $\alpha$-stable noise, i.e. the formal time derivative of the symmetric $\alpha$-stable motion \cite{janicki1994b}.
\bd{Note, that in the L\'evy flight (LF) scenario we do not take into consideration inertial effects, and similar to the Wiener process
we are dealing with an overdamped kind of motion. In contrast, in the L\'evy walk (LW) scheme, soon to be considered, we do include inertial effects
and finite propagation velocity. Hence one origin of the difference between both models stems from neglecting the inertia.}

Eq.~(\ref{eq:langevin}) is supplemented with the initial condition $x(0)=x_0$.
Accordingly, the stochastic process $\{X(t), t\geqslant 0\}$ governed by Eq.~(\ref{eq:langevin}) has increments
\begin{equation}
 \Delta x=x(t+\Delta t)-x(t)=\Delta t^{1/\alpha} \zeta_t
 \label{eq:discretization}
\end{equation}
distributed according to the symmetric $\alpha$-stable density with the scale parameter depending on the discretization time step $\Delta t$.
The discretization procedure assures that for a free particle with an arbitrary $\Delta t$, time dependent densities do not depend on the discretization time step but on $t$ (and remaining parameters) only.
In Eq.~(\ref{eq:discretization}) $\zeta_t$ represents independent, identically distributed ({\it i.i.d}) random variables following the symmetric $\alpha$-stable density \cite{janicki1994,janicki1996} with the characteristic function $\phi(k)$
\begin{equation}
 \phi(k)=\exp\left[ -\sigma_0^\alpha |k|^\alpha \right],
 \label{eq:fcharakt}
\end{equation}
where $\sigma_0>0$ is the scale parameter. The stability index $\alpha$ describes asymptotic (large $\Delta x$) behavior of the jump length density:
\begin{equation}
 p_\alpha(\Delta x;\sigma_0) \sim \frac{\sigma_0^\alpha \Delta t \sin \frac{\pi\alpha}{2} \Gamma(\alpha+1) /\pi}{|\Delta x|^{\alpha+1}}.
 \label{eq:asymptotics}
\end{equation}
Note that the parameter $\sigma_0$ scales the overall distribution width, hence its role is similar to a standard deviation for distributions with a finite second moment.
From Eq.~(\ref{eq:langevin}) and arithmetic properties of $\alpha$-stable densities \cite{janicki1996,samorodnitsky1994} it follows that the process $\{X(t), t\geqslant 0\}$ is distributed according to the $\alpha$-stable density with the time dependent parameter $\sigma(t)$
\begin{equation}
 \sigma(t)=\sigma_0 t^{1/\alpha},
 \label{eq:sigma}
\end{equation}
where $\sigma_0$ is a fixed scale parameter associated with the underlying $\alpha$-stable white noise in Eq.~(\ref{eq:langevin}).
Consequently, its asymptotics is described by Eq.~(\ref{eq:asymptotics}) after substitution of $\Delta t$ with $t$ and $\Delta x$ with $x$.

The Langevin equation~(\ref{eq:langevin}) can be associated with the space-fractional Smoluchowski-Fokker-Planck equation
\begin{equation}
 \frac{\partial P(x,t|x_0,0)}{\partial t} = \sigma_0^\alpha \frac{\partial^\alpha P(x,t|x_0,0)}{\partial |x|^\alpha} = K_\alpha \frac{\partial^\alpha P(x,t|x_0,0)}{\partial |x|^\alpha},
 \label{eq:ffpe}
\end{equation}
which describes evolution of the probability density function (PDF) of finding a random walker in the vicinity of $x$ at time $t$ with the initial condition $P(x,0|x_0,0)=\delta(x-x_0)$. The fractional operator
$\frac{\partial^\alpha}{\partial |x|^\alpha}$ stands for the fractional Riesz-Weil derivative, defined by the Fourier transform \cite{podlubny1998,samko1993}
$
 \mathcal{F}_k( \frac{\partial^\alpha f(x)}{\partial |x|^\alpha} )=-|k|^\alpha \mathcal{F}_k(f(x)).
$
In what follows, we interpret $K_\alpha=\sigma_0^\alpha$ in Eq.~(\ref{eq:ffpe}) as the generalized diffusion constant.

In the next subsections the main scope is to investigate interrelationship between formulation of boundary conditions for L\'evy flights and properties of escape kinetics and stationary states.
More precisely, in order, to assess various formulation of boundary conditions we explore two scenarios of escape kinetics from finite intervals (\textit{a}) restricted by two absorbing boundaries and (\textit{b}) restricted by reflecting and absorbing boundaries. We compare exact results (when applicable) with numerically estimated mean first passage times.
Moreover, for a finite interval restricted by two reflecting boundaries we verify if numerically constructed stationary densities agree with analytical predictions.
Finally, we study the various finite interval setups not only for L\'evy flights but also for L\'evy walks, which contrary to L\'evy flights have continuous trajectories and finite propagation velocity.
Comparison between behavior of the two models (LF and LW), with respect to a class of important observables, define the main scope of current research.
Depending on the value of the stability index $\alpha$ and observable type we find similarities but also large deviations between the two models.

\subsection{First escape problem\label{sec:fep}}

In the presence of boundary conditions imposed for Eq.~(\ref{eq:ffpe}) the translational invariance is broken and the resulting evolution equation for PDFs becomes a non-trivial integro-differential equation with non-local boundary conditions \cite{dybiec2006,zoia2007}. In order to avoid the problem, analysis of first escape is studied here by use of the Langevin methods, for which -- contrary to the methods based on the fractional Smoluchowski-Fokker-Planck equation -- the implementation of boundary conditions is significantly simpler, although not fully resolved.

\subsubsection{Absorbing boundaries at both ends\label{subsec:aa}}

\begin{figure}[!ht]
\includegraphics[angle=0, width=0.7\columnwidth]{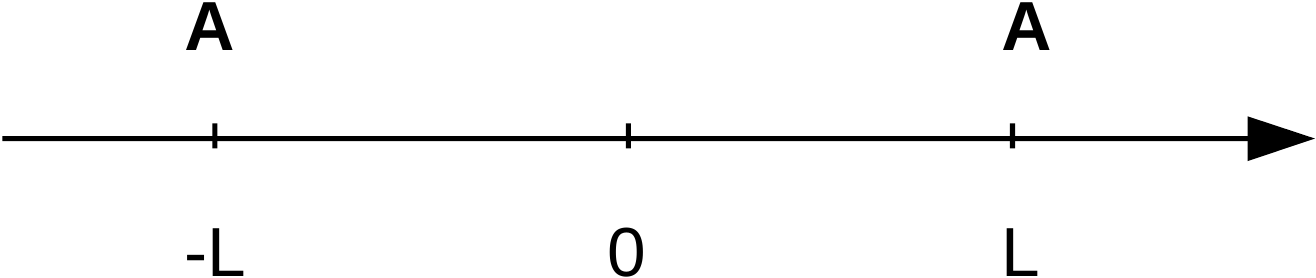}\\
\includegraphics[angle=0, width=\columnwidth]{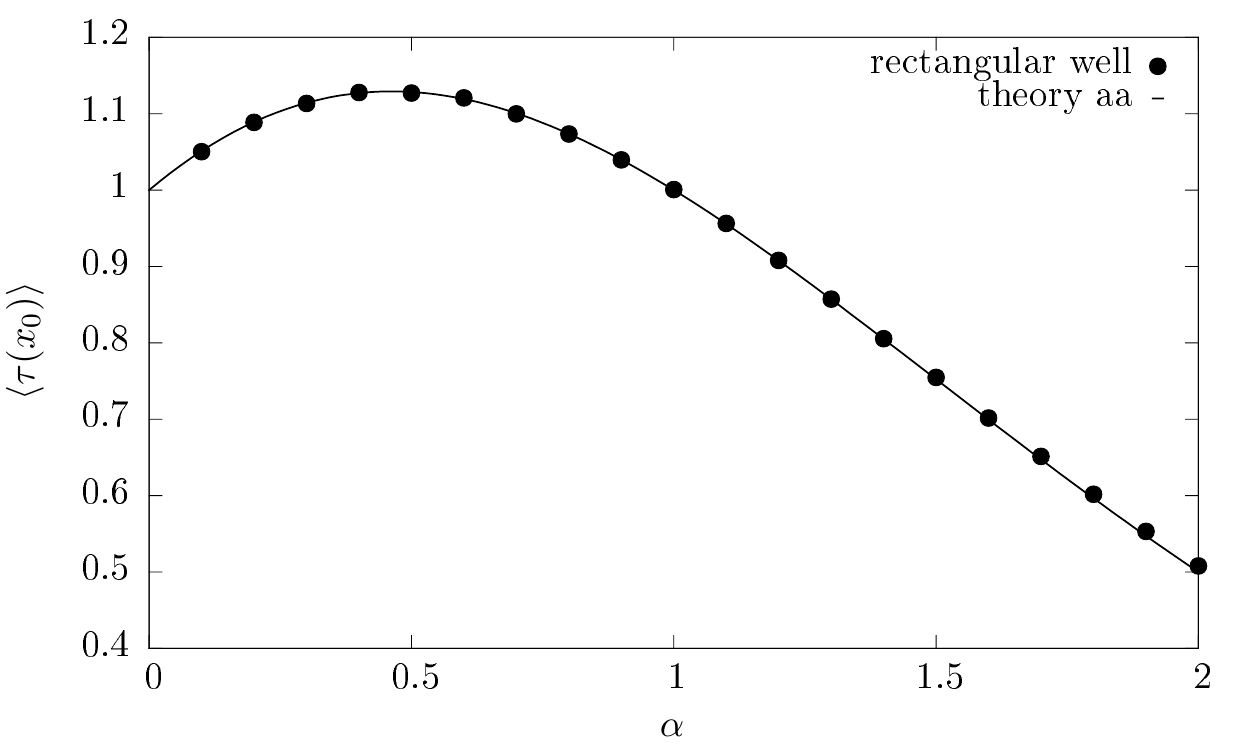}
\caption{Mean first passage time $\langle \tau (x_0) \rangle $ for absorbing-absorbing setup.
Points represent computer simulations for a finite interval ($\bullet$), see Eq.~(\ref{eq:langevin}).
Simulations parameters: interval half-width $L=1$, scale parameter $\sigma_0=1$, initial condition $x_0=0$, integration time step $\Delta t=10^{-4}$ and number of repetitions $N=10^6$.
Solid line presents the theoretical formula given by Eq.~(\ref{eq:general-mfpt}).}
\label{fig:aa}
\end{figure}

We consider a first escape problem from the $[-L,L]$ interval with both boundaries being absorbing, see Fig.~\ref{fig:aa}.
The evolution of $x(t)$ is determined by the Langevin Eq.~(\ref{eq:langevin}) and
the first passage time $\tau (x_0)$ ($|x_0| \leqslant L$) is defined as
\begin{equation}
 \tau (x_0)= \min\{t>0 \;\;:\;\; x(0)=x_0 \mbox{ and } |x(t)| \geqslant L \}.
 \label{eq:mfpt-trajectory}
\end{equation}
In this case, the formula for the mean first passage time (MFPT, $\langle \tau (x_0) \rangle$) reads \cite{getoor1961,zoia2007}
\begin{equation}
\langle \tau (x_0) \rangle
=\frac{1}{\Gamma(1+\alpha)} \frac{(L^2-|x_0|^2)^{\alpha/2}}{\sigma_0^\alpha}
\label{eq:general-mfpt}
\end{equation}
demonstrating that the MFPT asymptotically scales as $(L/\sigma_0)^\alpha$, what is especially well visible for $x_0=0$ when such a scaling is recorded for the whole $L$ range.
Note that, similar formula can be also found in \cite{buldyrev2001a}. 
Eq.~(\ref{eq:general-mfpt}) can be averaged over initial conditions. For example assuming that $x_0$ is uniformly distributed over $[-L,L]$ the mean exit time reads
\begin{equation}
 \langle \tau \rangle = \frac{\pi }{2^{1+\alpha} \Gamma\left[ (1+\alpha)/2 \right] \Gamma\left[ (3+\alpha)/2 \right]} \frac{L^\alpha}{\sigma_0^\alpha},
\end{equation}
which has exactly the same $(L/\sigma_0)^\alpha$ dependence like Eq.~(\ref{eq:general-mfpt}) for $x_0=0$. Therefore, in forthcoming considerations the fixed initial $x_0=0$ condition is used.

In numerical simulations of the corresponding Langevin equation the absorption condition is realized by assuming that the whole exterior of the prescribed interval is absorbing, i.e. each time when a trajectory crosses the absorbing boundary, it is removed and the first passage time is recorded.
Results of numerical simulations and formula~(\ref{eq:general-mfpt}) are presented in Fig.~\ref{fig:aa} showing perfect agreement with the theoretical curve.

In Fig.~\ref{fig:aa}, the scale parameter and the interval half-width have been arbitrary preset to $\sigma_0=1$, $L=1$ and the initial condition to $x_0=0$. Consequently, the $\langle \tau (x_0) \rangle (\alpha)$ curve attains one of its possible forms. In more general cases of $\sigma_0 \neq 1$ and $L \neq 1$, or more precisely the ratio $L/\sigma_0 \neq 1$, this curve can be of very different type, see \cite{szczepaniec2015escape}.

\begin{figure}[!ht]
\includegraphics[angle=0, width=\columnwidth]{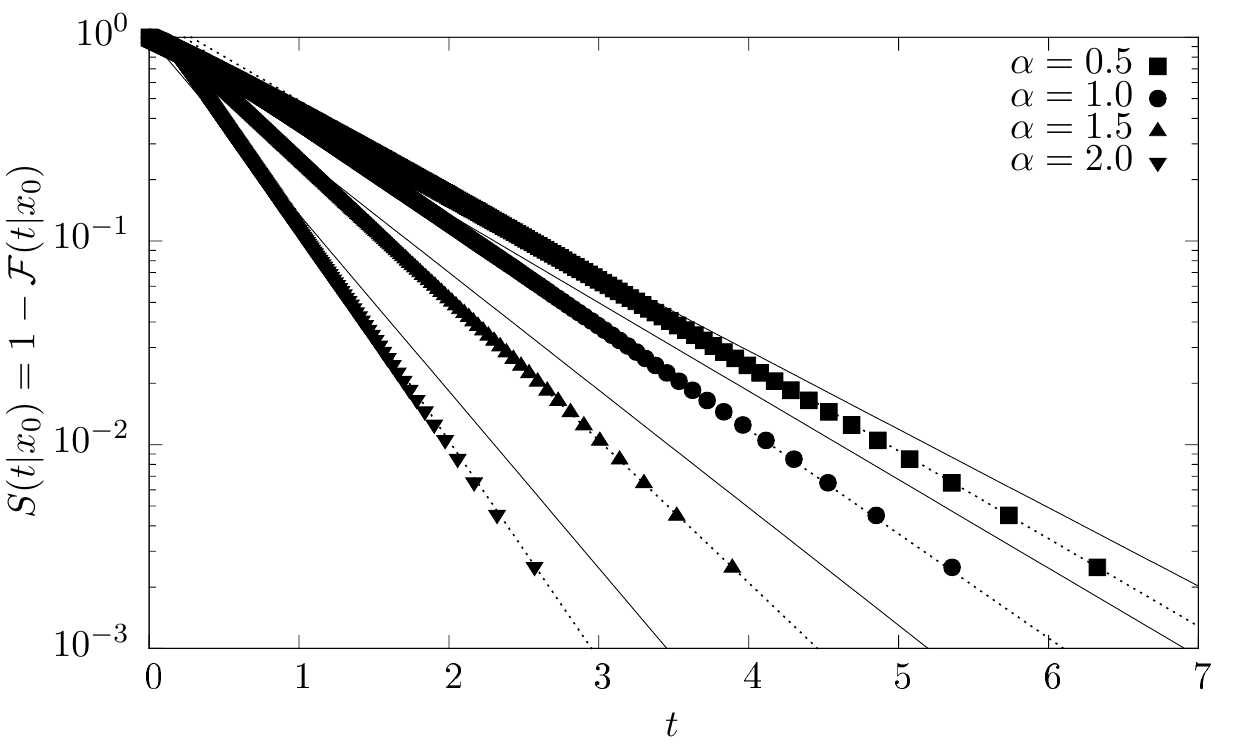}\\
\caption{Survival probabilities $S(t|x_0)$ corresponding to Fig.~\ref{fig:aa}, i.e. $x_0=0$.
\textcolor{black}{Solid lines present $\exp[-t/\langle \tau (x_0) \rangle]$ approximation to survival probabilities.}
Dotted lines depict $S(t|x_0) \approx \exp\left[-\lambda_1^{(\alpha)} t \right]$ approximation, see Eq.~(\ref{eq:eigenkwasnicki}).
}
\label{fig:aa-fpt}
\end{figure}

The cumulative distribution of first passage times for this problem defines the survival probability
\begin{equation}
 S(t|x_0)=1-\mathcal{F}(t|x_0)=1-\int_0^t p(s|x_0) ds,
 \label{eq:surv}
 \end{equation}
derived with the corresponding PDF $p(t|x_0)$ and depicted in Fig.~\ref{fig:aa-fpt}. Clearly,
the survival probability denotes the probability that a process starting at $x(0)=x_0=0$ has not reached or crossed up to time $t$ the levels $\pm L$.
Note that, by construction, the process described by Eq.~(\ref{eq:langevin}) is Markovian, which remains in line with observation of
 exponential asymptotics in Fig.~\ref{fig:aa-fpt}. The behavior is well documented in simulations of L\'evy flights \cite{dybiec2006,dybiec2014estimation} and can be inferred by estimation of lower and upper bounds \cite{katzav2008spectrumfractional,chen2010heat,kim2015first} for tails of $S(t|x_0)$
 or from the master equation \cite{drysdale1998,buldyrev2001a,araujo2016levy}. It can be also deduced by separation of variables \cite{zoia2007,dybiec2014estimation}
\begin{equation}
 S(t|x_0) = \sum_{i=1}^\infty c_i(x_0) \exp\left[ -\lambda_i^{(\alpha)} t \right],
 \label{eq:sumeigen}
\end{equation}
with $\lambda_i^{(\alpha)}$ denoting eigenvalues of the fractional Laplacian on bounded domains \cite{kwasnicki2012eigenvalues}.
Accordingly, the decay of the survival probability $S(t|x_0)$ is determined by the smallest eigenvalue of the fractional Laplacian \cite{zoia2007,katzav2008spectrumfractional,kwasnicki2012eigenvalues,dybiec2014estimation}, prompting a long time approximation
\begin{equation}
 S(t|x_0) \approx \exp\left[-\lambda_1^{(\alpha)} t\right].
 \label{eq:singleexpapproximation}
\end{equation}
The smallest eigenvalue $\lambda_1^{(\alpha)}$ can be calculated according to \cite[Eq.~(2)]{kwasnicki2012eigenvalues} which for $L=1$ reads
\begin{equation}
 \lambda_n^{(\alpha)}=\left[ \frac{n\pi}{2} + \frac{(2-\alpha)\pi}{8} \right]^\alpha.
 \label{eq:eigenkwasnicki}
\end{equation}
\textcolor{black}{
Eq.~(\ref{eq:singleexpapproximation}) along with properties of the survival probability, i.e. $\langle \tau (x_0)\rangle = \int_0^\infty S(t|x_0)dt$, suggest another possible approximation to the survival probability
\begin{equation}
 S(t|x_0) \approx \exp\left[-t/\langle \tau (x_0)\rangle \right],
 \label{eq:approximation}
\end{equation}
where $\langle \tau (x_0) \rangle$ is the mean first passage time given by Eq.~(\ref{eq:general-mfpt}).
}

\textcolor{black}{
Figure~\ref{fig:aa-fpt} compares both approximations, see Eqs.~(\ref{eq:singleexpapproximation}) and (\ref{eq:approximation}).
}
\textcolor{black}{
Solid lines represent Eq.~(\ref{eq:approximation}): L\'evy motion on a confined interval between two absorbing boundaries decays with the steepness parameter depending on the stability index $\alpha$. At the same time, with increasing $\alpha$ the deviations from a single exponential approximation given by Eq.~(\ref{eq:approximation})
become more pronounced, as more and more terms from Eq.~(\ref{eq:sumeigen}) have to be retained \cite{katzav2008spectrumfractional} in order to reconstruct an initial part of the survival probability. Therefore, approximation (\ref{eq:approximation}) does not reproduce correct decay rate of the survival probability.
}
\textcolor{black}{
Additional dotted lines in Fig.~\ref{fig:aa-fpt} depict single exponential, smallest eigenvalue approximation (\ref{eq:singleexpapproximation}), which does not work perfectly, but it predicts the right exponent characterizing the asymptotic slope of the survival probability.
}

\begin{figure}[!ht]
\includegraphics[angle=0, width=\columnwidth]{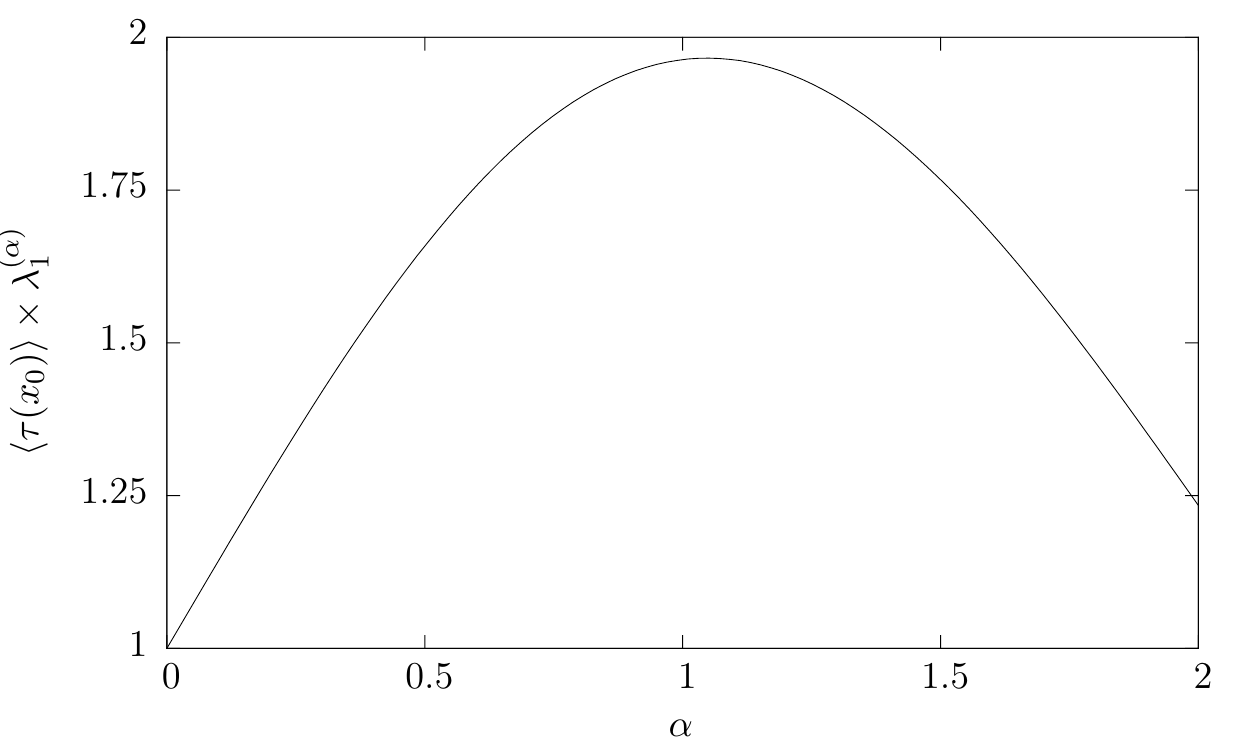}\\
\caption{\textcolor{black}{The ratio of the MFPT $\langle \tau (x_0) \rangle $ and MFPT approximation $1/\lambda_1^{(\alpha)}$, i.e. $\langle \tau (x_0) \rangle \times \lambda_1^{(\alpha)}$.
The interval half-width $L=1$ and the initial condition $x_0=0$.
}}
\label{fig:eigen}
\end{figure}

\textcolor{black}{
In particular, due to pedagogical reasons, approximations (\ref{eq:singleexpapproximation}) and (\ref{eq:approximation}) can be compared for $\alpha=2$. In such a case, the MFPT can be calculated from Eq.~(\ref{eq:general-mfpt})
\begin{equation}
 \langle \tau (0) \rangle = \frac{L^2}{2 \sigma_0^2}.
 \label{eq:mfptAlpha2}
\end{equation}
The smallest eigenvalue of the Laplacian is \cite{cox1965,dybiec2012fractional}
\begin{equation}
 \lambda_1^{(2)}= \frac{\pi^2\sigma_0^2}{4 L^2},
\end{equation}
leading to $\langle \tau (x_0) \rangle = 1/\lambda_1^{(2)}=4L^2/\pi^2 \sigma_0^2 \approx 0.405 L^2/\sigma_0$, which differs by $24\%$ from the exact value, see Eq.~(\ref{eq:mfptAlpha2}).
The quality of approximation (\ref{eq:approximation}) depends on the exact value of the stability index $\alpha$ what can be inferred from Fig.~\ref{fig:eigen}, which presents the ratio of the exact value of the MFPT, see Eq.~(\ref{eq:general-mfpt}), to its approximation $\langle \tau (x_0)\rangle \approx 1/\lambda_1^{(\alpha)}$. The dependence of $\langle \tau (x_0) \rangle  \times  \lambda_1^{(\alpha)}$ is a non monotonous function of the stability index $\alpha$ with the maximum located around $\alpha \approx 1$. For small $\alpha$ Eqs.~(\ref{eq:singleexpapproximation}) and (\ref{eq:approximation}) provide reasonable approximations to the survival probability.
Finally, it is worthy to underline, that quality of the approximation (\ref{eq:approximation}) becomes worse with the increasing number of spatial dimensions \cite{dybiec2015escape}.
}

\subsubsection{Reflecting-absorbing boundary conditions\label{subsec:ra}}

\begin{figure}[!ht]
\includegraphics[angle=0, width=0.7\columnwidth]{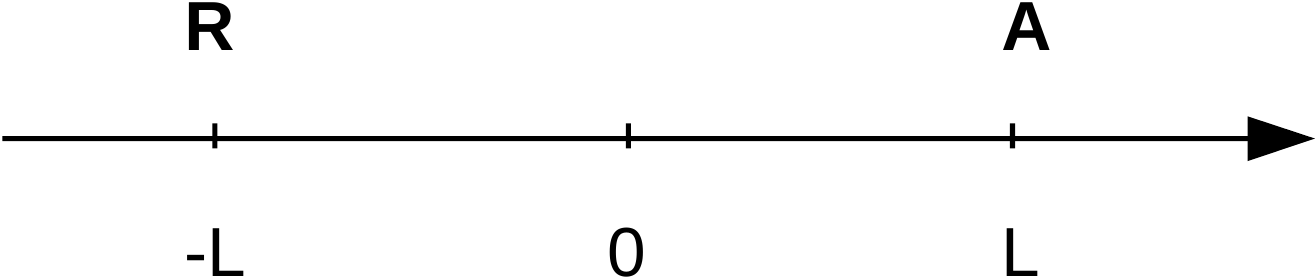}\\
\includegraphics[angle=0, width=\columnwidth]{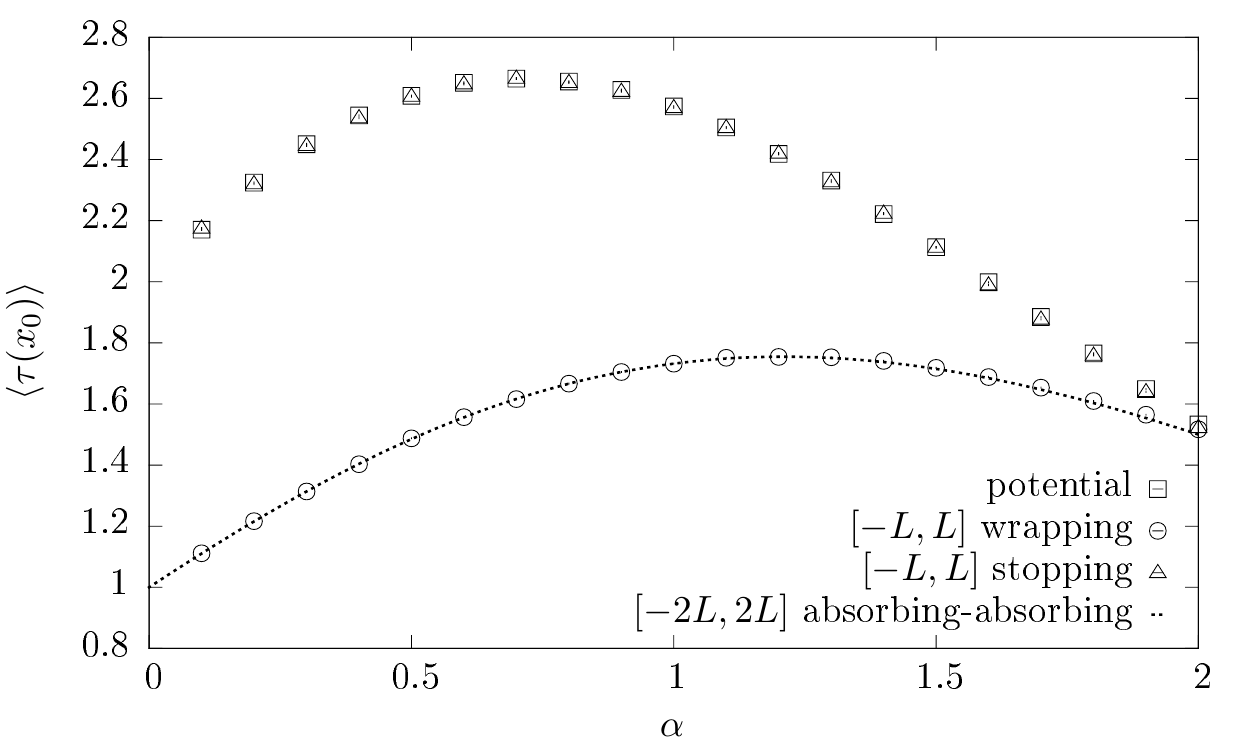}\\
\includegraphics[angle=0, width=\columnwidth]{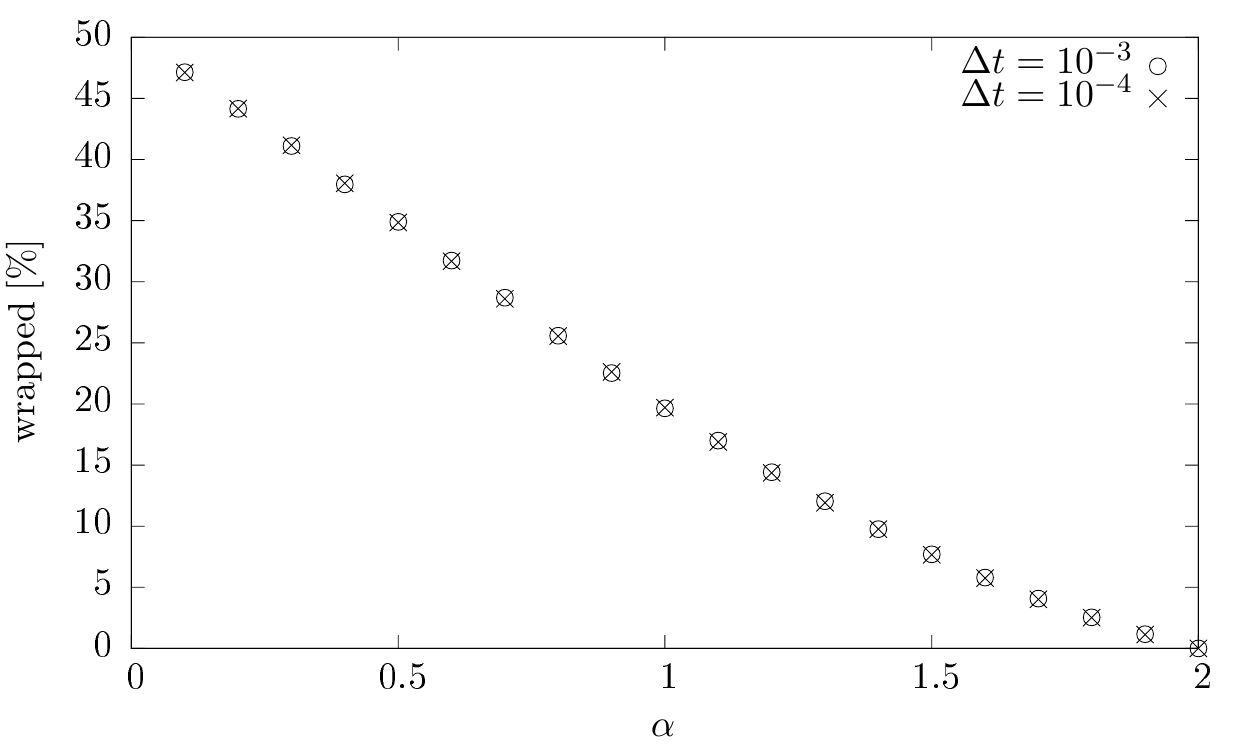}
\caption{Mean first passage time $\langle \tau (x_0) \rangle$ for reflecting-absorbing setup (middle panel).
Points represent computer simulations for infinite rectangular potential well $[-L,L]$ with reversing ($\circ$), rectangular potential well $[-L,L]$ with stopping ($\triangle$) and potential $V_{800}(x)$ ($\square$).
Short-dashed line presents formula for the MFPT (from the $[-2L,L]$ restricted by two absorbing boundaries) given by Eq.~(\ref{eq:general-mfpt}).
Bottom panel presents fraction of escape events due to wrapping of trajectories along the reflecting boundary located at $-L$.
Simulations parameters: $L=1$, $\sigma_0=1$, $x_0=0$, $\Delta t=10^{-4}$ and $N=10^6$.}
\label{fig:ra}
\end{figure}

Next, we consider a first escape problem from the $[-L,L]$ interval with reflecting (left) and absorbing (right) boundaries, see Fig.~\ref{fig:ra}.
The first passage time $\tau (x_0) $ ($|x_0|\leqslant L$) is then defined as
\begin{equation}
 \tau (x_0) = \min\{t>0 \;\;:\;\; x(0)=x_0 \mbox{ and } x(t) \geqslant L \}.
 \label{eq:mfpt-trajectory-r}
\end{equation}
Analogously, like in Sec.~\ref{subsec:aa} we use $x_0=0$.

Imposing a reflecting boundary at $x=-L$ requires some additional care in numerical simulations. Here we consider three different realizations of the reflecting condition:
\begin{itemize}
 \item[(\textit{i})] \textit{motion reversal}: a trajectory which ends at $x<-L$ is wrapped around the left boundary, i.e. $x \to-L+|x+L|$;
 \item[(\textit{ii})] \textit{motion stopping}: a trajectory which crosses $-L$ is paused at $-L+\varepsilon$, where $\varepsilon$ is a small and positive parameter. The point $-L+\varepsilon$ is used as starting point for a next jump;
 \item[(\textit{iii})] \textit{motion confined within a potential}: the reflecting boundary can be implemented by considering the motion in a bounding potential
 \begin{equation}
 \lim_{n\to\infty}V_n(x) = \lim_{n\to\infty}\frac{1}{n} \frac{|x|^n}{L^n},
 \label{eq:vn}
 \end{equation}
 which is described by the following Langevin equation
 \begin{equation}
 \frac{dx}{dt}=-V_n'(x)+\zeta_\alpha(t).
 \label{eq:langevin2}
 \end{equation}
 For $n\geqslant 2$, the potential~(\ref{eq:vn}) and its first derivative (force) are continuous.
 In the limit of $n\to\infty$, the potential $V_n(x)$ reduces to the infinite steep rectangular potential well with (reflecting) boundaries at $\pm L$.
\end{itemize}
In all above scenarios we use Langevin equations (\ref{eq:langevin}) or (\ref{eq:langevin2}) to simulate the L\'evy motion.
The absorbing boundary is executed in a standard way, i.e. whenever $x\geqslant L$, a particle becomes absorbed, both for a free particle motion and motion in the potential well~(\ref{eq:vn}).

As it can be inferred from the middle panel of Fig.~\ref{fig:ra}, the \textit{stopping} and \textit{potential} scenarios yield same values of estimated MFPT, conditioned that $n$ and $\varepsilon$ are sufficiently large and small, respectively.
In contrast, the implementation of the \textit{wrapping} method underestimates MFPT in comparison to the other two cases.
All scenarios are equivalent for $\alpha=2$ indicating that the source of discrepancy lies in discontinuity of trajectories for L\'evy motion with $\alpha<2$.
In fact, this conclusion can be drawn by a more accurate analysis of wrapping trajectories: the bottom panel of Fig.~\ref{fig:ra} presents the fraction of escape events (\textit {wrapping scenario}) in which a particle staring at $x_0=0$ escaped from the $[-1,1]$ interval by a single long jump to the left and has been reversed around the reflecting boundary. This fraction decreases with increase of the stability index $\alpha$ and tends to be arbitrarily small in the Gaussian limit ($\alpha\to 2$), when the trajectory $x(t)$ becomes continuous.
In the opposite limit of $\alpha \to 0$ almost half of escape events are due to \textit{trajectory wrapping}, as might be expected for extremely wide and heavy tailed PDF of increments $\Delta x$.

Additionally to the analysis of escape events from the $[-L,L]$ interval (reflecting-absorbing), the escape from the $[-2L,2L]$ interval (absorbing-absorbing) has been investigated. Such a scenario gives the same results as \textit{wrapping}, since the escape from the reflecting-absorbing interval is equivalent to the escape from two times wider absorbing-absorbing interval, see central panel of Fig.~\ref{fig:ra} where
$\langle \tau (0) \rangle$ given by Eq.~(\ref{eq:general-mfpt}) for a system $[-2L , 2L]$ is depicted.
The theoretical formula perfectly matches the simulations on the half sized system with the reversing (wrapping) condition.
For $\alpha=2$ and various types of boundary conditions, this correspondence can be trivially read off from the analytic formula for the MFPT \cite{gardiner2009}.
In more general settings with $\alpha < 2$, the equivalence relies on infinite propagation of the trajectory $x(t)$: from any point $x$ the distance to the absorbing boundary is either $L-x$ or $x+3L$ (when wrapped along a reflecting boundary). The sum of these two distances is the only relevant model parameter and equals exactly the sum of distances to absorbing boundaries located at $\pm 2L$.

\begin{figure}[!ht]
\includegraphics[angle=0, width=\columnwidth]{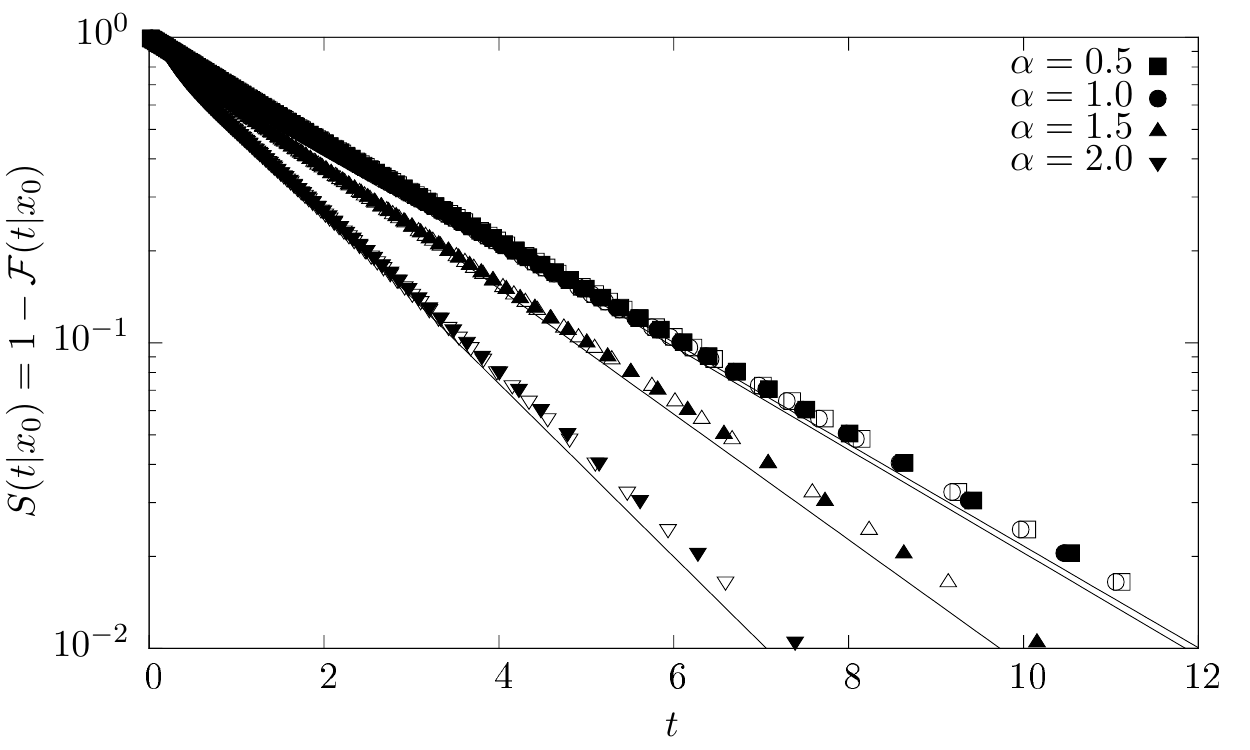}\\
\includegraphics[angle=0, width=\columnwidth]{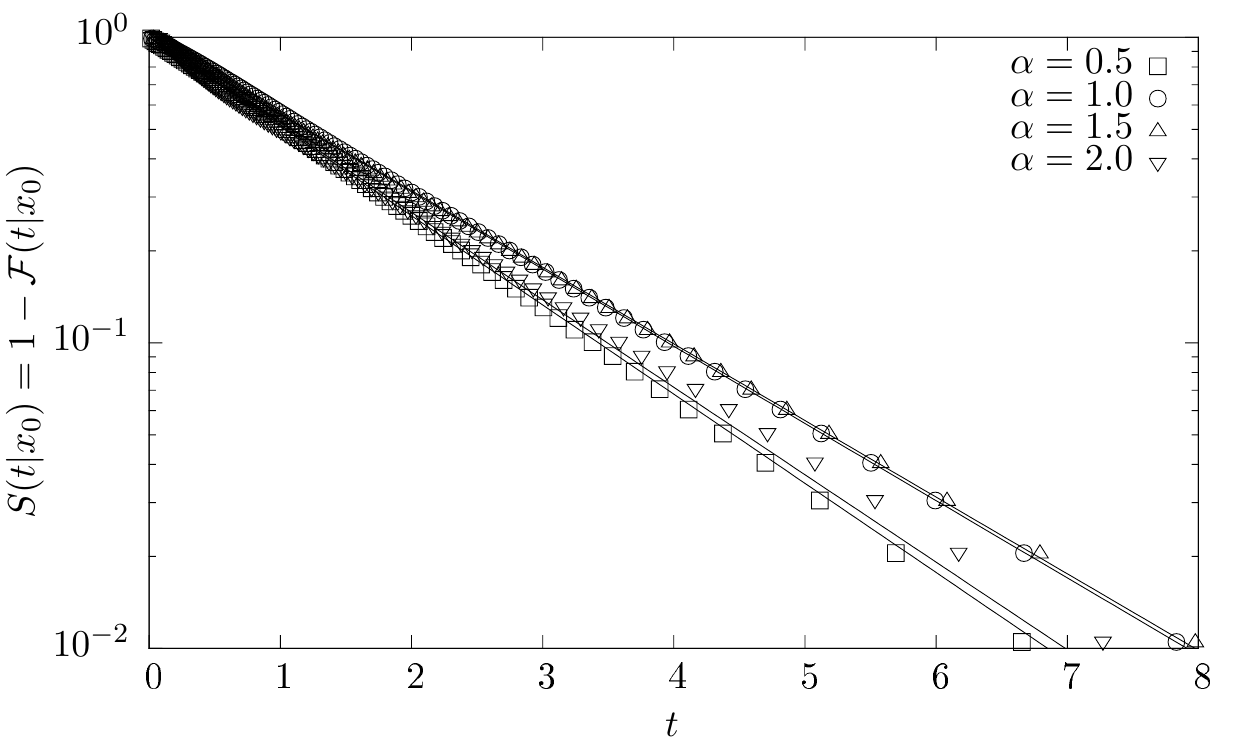}\\
\caption{
Survival probabilities $S(t|x_0)$ corresponding to Fig.~\ref{fig:ra}, i.e. $x_0=0$. In the top panel: filled points represent results for $V_{800}(x)$ potential while empty symbols correspond to infinite steep rectangular potential well with stopping.
The bottom panel presents results for infinite steep rectangular potential well with wrapping.
\textcolor{black}{
Solid lines present $S(t|x_0)\approx \exp\left[ -t/\langle \tau (x_0) \rangle \right]$, where $\langle \tau (x_0) \rangle$ is the MFPT from Fig.~\ref{fig:ra} corresponding to the appropriate scenario.
}
}
\label{fig:ra-fpt}
\end{figure}

Figure~\ref{fig:ra-fpt} presents survival probabilities $S(t|x_0)$ corresponding to motions analyzed in Fig.~\ref{fig:ra}.
The top panel presents results for motion in the potential $V_{800}(x)$ and in a rectangular potential well with stopping.
These two scenarios result in the same first passage time PDFs and, accordingly, in the same mean first passage times.
The bottom panel of Fig.~\ref{fig:ra-fpt} exemplifies results for the survival probability in the \textit{wrapping scenario} which leads to faster escape kinetics.
Analogously to the absorbing-absorbing setup, the first passage time densities display exponential asymptotics reflecting Markovian characteristics of the process $x(t)$, cf. Eqs.~(\ref{eq:langevin}) and (\ref{eq:langevin2}).

\textcolor{black}{
Solid lines in Fig.~\ref{fig:ra-fpt} represent $S(t|x_0) \approx \exp\left[ -t/\langle \tau (x_0) \rangle \right]$ approximation to the survival probability.
Analogously, like in Fig.~\ref{fig:aa-fpt} this kind of approximation works better for small values of the stability index $\alpha$. Therefore, the largest discrepancies are observed for $\alpha=2$. For the wrapping scenario discrepancies are smaller than for remaining scenarios, because wrapping of trajectories accelerates decay of the survival probability.
}

\subsection{Stationary states\label{subsection:stationary}}

\begin{figure}[!ht]
\includegraphics[angle=0, width=0.7\columnwidth]{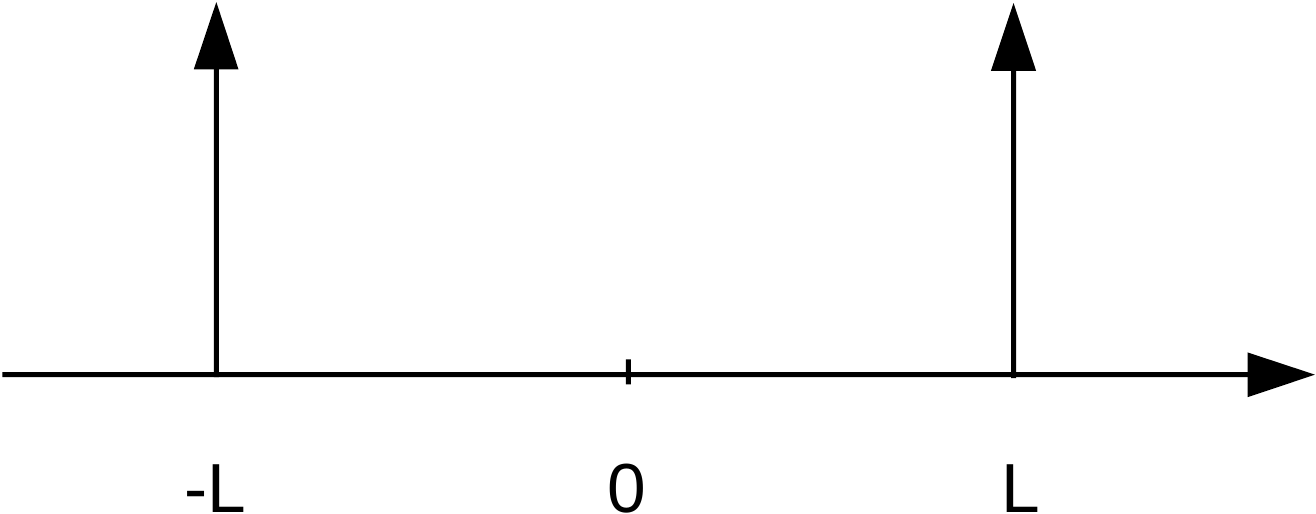}\\
\includegraphics[angle=0, width=\columnwidth]{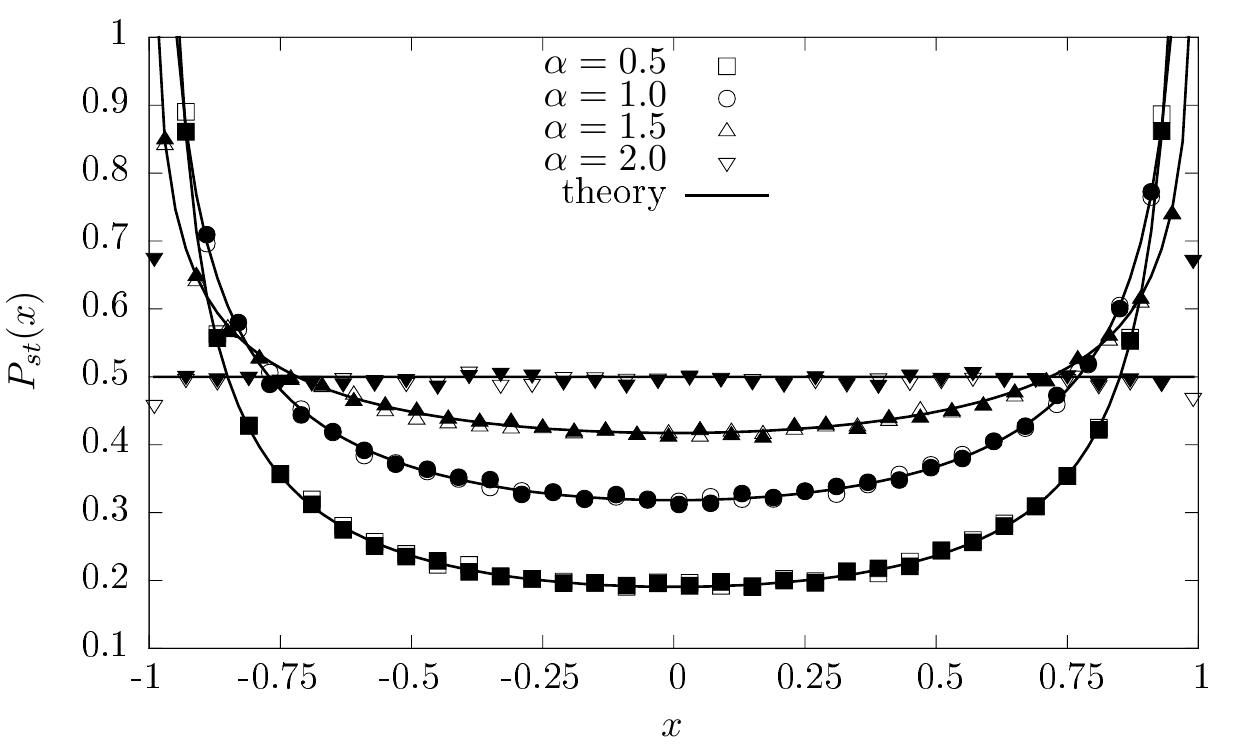}\\
\includegraphics[angle=0, width=\columnwidth]{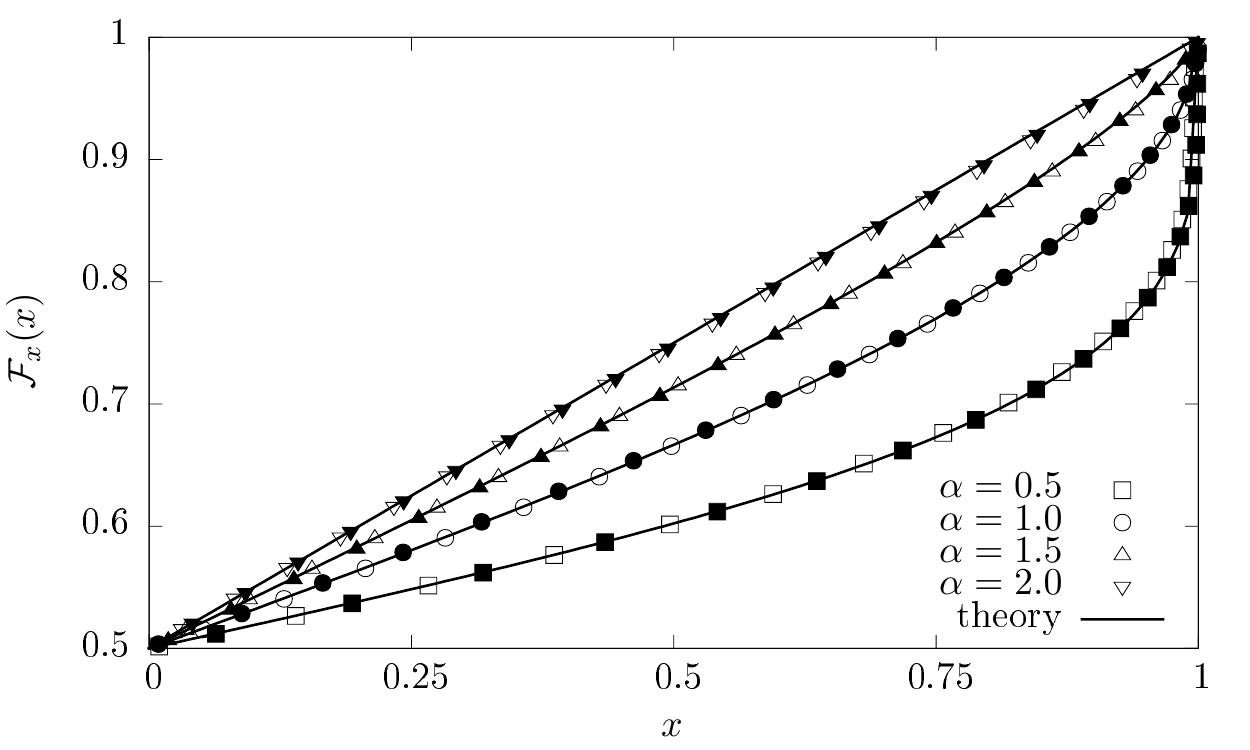}\\
\caption{The middle panel presents stationary states for infinitely deep rectangular potential well with the stopping scenario (empty points) and the potential $V_{800}(x)$, see Eq.~(\ref{eq:vn}), (filled points).
Solid lines present theoretical formula given by Eq.~(\ref{eq:pst-rw}.)
Bottom panel presents cumulative densities $\mathcal{F}_x(x)$ corresponding to histograms from the middle panel. Solid lines depict cumulative densities $\mathcal{F}_x(x)$ calculated from Eq.~(\ref{eq:pst-rw}).
In order to improve figure's clarity $\mathcal{F}_x(x)$ are plotted for $0\leqslant x \leqslant 1$ only.
Small discrepancies, in the middle panel, at $x \approx \pm 1$ are due to data binning.
}
\label{fig:his}
\end{figure}

In case of L\'evy flights in an infinitely deep rectangular well, a particle executing the motion becomes confined in a domain restricted by two impenetrable boundaries (cf. top panel of Fig.~\ref{fig:his}).
With these conditions, stationary states can be observed.
In what follows, we compare simulations of Eq.~(\ref{eq:langevin2})
in the limit of large $n$ with the results of simulated free diffusion bounded by the \textit{stopping} or \textit{wrapping} conditions at reflecting boundaries at $x=\pm L$.
The \textit{stopping} (empty points) and \textit{potential} (filled points) scenarios produce stationary PDFs displayed in Fig.~\ref{fig:his}, in full agreement with analytical result derived by Denisov et al. \cite{denisov2008}:
\begin{equation}
P_{st}(x)=\frac{\Gamma(\alpha)(2L)^{1-\alpha}(L^2-x^2)^{\alpha/2-1}}{\Gamma^{2}(\alpha/2)}.
\label{eq:pst-rw}
\end{equation}
Bottom panel of Fig.~\ref{fig:his} depicts cumulative densities $\mathcal{F}_x(x)=\int_{-L}^x P_{st}(x')dx'$ corresponding to histograms presented in Fig.~\ref{fig:his} and calculated (solid lines) from Eq.~(\ref{eq:pst-rw}).
For clarity of the presentation, the cumulative densities have been plotted for $0 \leqslant x \leqslant L$ with $L=1$ only.

\subsection{Mean Residence time\label{subsec:mrt}}

It is instructive to compare results of the above MFPT analysis with the mean residence time (MRT).
MRT represents the average time that a freely diffusing particle, moving on $-\infty<x(t)<\infty$ resides in a given region (say, in the interval $[-L,L]$) in a measurement process of duration $t$. Consequently, the MRT is always shorter than the measurement time $t$.
Moreover, due to possible multiple returns to the $[-L,L]$ interval MRT can be larger than MFPT.
We will consider two limits, when measurement time is infinite, and also the time dependence of the MRT.
We start with evaluating the probability to find a particle in the interval
$[-L,L]$ at the time $t$, provided it has started from some internal
point $x_0\in\left(-L,L\right)$
\begin{equation}
\mathrm{Pr}\left( t,x_0\right) =\int_{-L}^{L}P\left(\left.
x,t\right| x_0,0\right) dx.\label{G-02}
\end{equation}
The probability density of transitions $P\left( \left. x,t\right|
x_{0},0\right)$ in Eq.~(\ref{G-02}) is the solution of
Eq.~(\ref{eq:ffpe}) with the initial condition
\begin{equation}
P\left(x,0\right) = \delta\left( x-x_{0}\right).\label{G-03}
\end{equation}

If we make the Fourier transform of Eq.~(\ref{eq:ffpe}) and introduce
the characteristic function which is defined as
\begin{equation}
\vartheta \left( k,t\right) =\int_{-\infty }^{\infty}P\left(
x,t\right) e^{ikx}\,dx \label{G-04}
\end{equation}
we arrive at
\begin{equation}
\frac{\partial\vartheta }{\partial t} = -\sigma_0^{\alpha}\left\vert
k\right\vert ^{\alpha }\vartheta .\label{G-05}
\end{equation}
From Eqs.~(\ref{G-03}) and (\ref{G-04}) we find the initial
condition for Eq.~(\ref{G-05})
\begin{equation}
\vartheta \left(k,0\right) = e^{ikx_0}.\label{G-06}
\end{equation}
The exact solution of Eq.~(\ref{G-05}) with the initial
condition~(\ref{G-06}) reads
\begin{equation}
\vartheta \left(k,t\right) = \exp\left[ikx_0 -
\sigma_0^{\alpha}\left\vert k\right\vert
^{\alpha}t\right],\label{G-07}
\end{equation}
which is the $\alpha$-stable density with the same stability index $\alpha$ like the underlying noise and the time dependent scale parameter $\sigma_0 t^{1/\alpha}$, compare Eqs.~(\ref{G-07}) and (\ref{eq:fcharakt}).
A non-zero initial condition introduce additional shift to the time dependent density, which is centered at $x_0$.

Using the inverse Fourier transform we obtain from Eq.~(\ref{G-07})
\begin{equation}
P\left(\left. x,t\right| x_0,0\right) = \frac{1}{2\pi}\int_{-\infty
}^{\infty}e^{-ik\left(x-x_0\right) - \sigma_0^{\alpha}\left\vert
k\right\vert ^{\alpha}t}dk.\label{G-08}
\end{equation}
Substituting Eq.~(\ref{G-08}) in Eq.~(\ref{G-02}) and integrating
over $x$ we arrive at
\begin{equation}
\mathrm{Pr}\left( t,x_0\right)
=\frac{2}{\pi}\int_{0}^{\infty}\frac{\sin{kL\cos{kx_0}}}{k}\,e^{-\sigma_0^{\alpha}k^{\alpha}t}\,dk.\label{G-09}
\end{equation}

According to the definition, MRT $\langle T(x_0) \rangle$ in the interval
$[-L,L]$ can be calculated as
\begin{equation}
\langle T\left(x_{0}\right) \rangle = \int_{0}^{\infty}\mathrm{Pr}\left(
t,x_{0}\right) dt.\label{G-10}
\end{equation}
This MRT is the mean residence time in the limit where the total measurement time is large, below we will obtain the time dependence of the MRT.
Substituting Eq.~(\ref{G-09}) in Eq.~(\ref{G-10}) and integrating
over $t$ we get
\begin{equation}
\langle T\left(x_{0}\right) \rangle =
\frac{1}{\pi\sigma_0^{\alpha}}\int_{0}^{\infty}\frac{\sin{k\left(L+x_0\right)}+\sin{k\left(L-x_0\right)}}{k^{1+\alpha}}\,dk.\label{G-11}
\end{equation}
As seen from Eq.~(\ref{G-11}), the integral diverges for $1
\leqslant\alpha <2$, i.e. MRT goes to infinity as for
standard Brownian diffusion ($\alpha =2$).
Divergence of the mean residence time is the consequence of multiple returns to the $[-L,L]$ interval.
In the case $0 <\alpha <1$
Eq.~(\ref{G-11}) yields
\begin{equation}
\langle T\left(x_{0}\right) \rangle =
\frac{\Gamma\left(1-\alpha\right)}{\pi\sigma_0^{\alpha}\alpha}\left[\left(L+x_0\right)^{\alpha}+\left(L-x_0\right)^{\alpha}\right]\sin{\frac{\pi\alpha}{2}}\,,\label{G-12}
\end{equation}
where $\Gamma(x)$ is the Euler gamma function. In particular, for $x_0 =0$
from Eq.~(\ref{G-12}) we have
\begin{equation}
\langle T(0) \rangle =
\frac{2\,\Gamma\left(1-\alpha\right)}{\pi\alpha}\left(\frac{L}{\sigma_0}\right)^{\alpha}\sin{\frac{\pi\alpha}{2}}\,.\label{G-13}
\end{equation}
MRT given by Eqs.~(\ref{G-12}) and (\ref{G-13}) is finite, but larger than the MFPT.
In the limit of $\alpha\to 0$, the mean residence time in Eq.~(\ref{G-13}) tends to $1$, which is the mean first passage time from the $[-L,L]$ interval given by Eq.~(\ref{eq:general-mfpt}).
Equality of MRT and MFPT signals that the particle after escaping the domain $[-L,L]$, does not return to the interval any more.
This is in accordance with properties of L\'evy flights, which for $\alpha<1$ are transient \cite{blumenthal1961}, i.e. for LF with $\alpha<1$ starting outside a finite interval, there is nonzero probability of not visiting this interval at all. In the limit of $\alpha \to 0$ this probability tends to 1.
In the intermediate range of $0 < \alpha < 1$, returns to the $[-L,L]$ interval are possible, but their probability is smaller than 1.
Therefore, MFPT is smaller than (finite) MRT.

The long time asymptotics of the probability $\mathrm{Pr}\left(
t,x_0\right)$ can be estimated from Eq.~(\ref{G-09}). In the limit $t\rightarrow
\infty$ the function
$\exp\left[-\sigma_0^{\alpha}k^{\alpha}t\right]$ under the integral
becomes very narrow near the point $k=0$ and we can approximately
estimate this integral. As a result, we arrive at
\begin{equation}
\mathrm{Pr}\left( t,x_0\right) \sim
\frac{2\,\Gamma\left(1/\alpha\right)L}{\pi\sigma_0\alpha\,
t^{1/\alpha}}\,,\quad t\rightarrow\infty.\label{G-14}
\end{equation}
As seen from Eq.~(\ref{G-14}) the integral~(\ref{G-10}) diverges for
$\alpha \geqslant 1$.

It is interesting to compare the result~(\ref{G-14}) with a
 free Brownian diffusion described by the following
Smoluchowski-Fokker-Planck equation
\begin{equation}
\frac{\partial P}{\partial t}= K_2\,\frac{\partial^{2}P}{\partial
x^{2}}\,.\label{G-15}
\end{equation}
The solution of Eq.~(\ref{G-15}) with the initial
condition~(\ref{G-03}) is well-known and has the form
\begin{equation}
P\left(\left. x,t\right| x_0,0\right) = \frac{1}{2\sqrt{\pi
Dt}}\exp\left[-\frac{\left(x-x_0\right)^2}{4Dt}\right].\label{G-16}
\end{equation}
Substituting Eq.~(\ref{G-16}) in Eq.~(\ref{G-02}) in the limit
$t\rightarrow \infty$ we approximately have
\begin{equation}
\mathrm{Pr}\left( t,x_0\right) \sim \frac{L}{\sqrt{\pi Dt}}\,,\quad
t\rightarrow\infty.\label{G-17}
\end{equation}
The asymptotics~(\ref{G-14}) transforms into asymptotics~(\ref{G-17})
when $\alpha =2$ because of $\Gamma(1/2)=\sqrt{\pi}$ and $\sigma_0 =
\sqrt{K_2}$.

Equations~(\ref{G-11}) and~(\ref{G-13}) provide formulas for the mean residence time when $\alpha<1$ and for infinitely long measurement times.
Depending on the observation time $t$, the amount of time spent in the $[-L,L]$ interval changes.
Therefore, a related question is to estimate how the mean residence (occupation) time grows with the measurement time $t$.
The scaling of the average occupation time as a function of the measurement time can be calculated by general properties of L\'evy flights. By the definition, the occupation time $T$ is
\begin{equation}
 T(x_0)=\int_0^t \Theta(x(t'))dt',
\end{equation}
where $x(t)$ is a random walker location, $t$ is the measurement time and $\Theta(\cdot)$ is the characteristic function
\begin{equation}
 \Theta(x(t))=\left\{
 \begin{array}{cl}
 1 & -L \leqslant x(t) \leqslant L \\
 0 & \mbox{otherwise} \\
 \end{array}
 \right..
\end{equation}
The average residence time $\langle T (x_0) \rangle$ is (for simplicity $x_0=0$)
\begin{equation}
 \langle T (0) \rangle = \int_0^t \langle \Theta(x(t')) \rangle dt'
 = \int_0^t dt' \int_{-L}^L P(x,t'|0,0) dx,
 \label{eq:fractiondef}
\end{equation}
where $P(x,t|0,0)$ is
\begin{equation}
P\left(\left. x,t\right| 0,0\right) = \frac{1}{2\pi}\int_{-\infty
}^{\infty}e^{-ik x - \sigma_0^{\alpha}\left\vert
k\right\vert ^{\alpha}t}dk,
\end{equation}
see Eq.~(\ref{G-08}).

For infinite measurement time ($t\to\infty$) and $0<\alpha<1$ the integral in Eq.~(\ref{eq:fractiondef}) is convergent
\begin{equation}
 \lim_{t\to\infty}\langle T (0)\rangle
 = \frac{2}{\pi\sigma_0^\alpha} \int_0^\infty \frac{\sin kL}{k^{\alpha+1}}dk
 =\frac{2}{\pi}\left( \frac{L}{\sigma_0} \right)^{\alpha} \int_0^\infty \frac{\sin y}{y^{\alpha+1}}dy
\end{equation}
and is equal to Eq.~(\ref{G-13}) representing the fact that for $\alpha<1$ a random walker spends a constant amount of time in the $[-L,L]$ interval, what is the consequence of already discussed transient character of $\alpha$-stable motions with $\alpha<1$, \cite{dybiec2016jpa}.

For finite measurement time $t$ and any $0 < \alpha \leqslant 2$
\begin{eqnarray}
 \langle T (0) \rangle & = & \frac{1}{\pi}\int_{-\infty}^\infty \frac{\sin kL}{k} \frac{1-e^{-|k|^\alpha\sigma_0^\alpha t}}{|k|^\alpha\sigma_0^\alpha} dk \\ \nonumber
 & = & \frac{2}{\pi\sigma_0^\alpha}\int_{0}^\infty \frac{\sin kL}{k^{1+\alpha}} \left[ 1-e^{-|k|^\alpha\sigma_0^\alpha t} \right] dk.
 \label{eq:genform}
 \end{eqnarray}
 Eq.~(\ref{eq:genform}) can be approximated for short and large measurement times.
 For the short measurement time one has
\begin{eqnarray}
 \langle T (0) \rangle & \approx & \frac{2}{\pi\sigma_0^\alpha}\int_{0}^\infty \frac{\sin kL}{k} \sigma_0^\alpha t dk \\ \nonumber
 & = & \frac{2t}{\pi} \int_0^\infty \frac{\sin y}{y} dy = t.
 \end{eqnarray}
 In order to find large measurement time asymptotics, the general formula given by Eq.~(\ref{eq:genform}) needs to be rewritten as
 \begin{eqnarray}
 \langle T (0) \rangle & = & \frac{2}{\pi} t \int_0^\infty \frac{\sin \kappa \xi}{\kappa^{1+\alpha}} \left[ 1-e^{-\kappa^\alpha} \right]d\kappa,
 \label{eq:genform2}
 \end{eqnarray}
 where $\xi=L/(\sigma_0 t^{1/\alpha})$ and $\kappa^\alpha=k^\alpha \sigma_0^\alpha t$.
From Eq.~(\ref{eq:genform2}) one gets the large $t$ (small $\xi$) asymptotics
\begin{eqnarray}
 \langle T (0) \rangle & \approx & \frac{2}{\pi} \xi \int_0^\infty \frac{1-e^{-\kappa^\alpha}}{\kappa^\alpha}d\kappa \\ \nonumber
 & = & \frac{2}{\pi} \frac{L}{\sigma_0} \frac{\Gamma(1/\alpha)}{\alpha-1} t^{1-1/\alpha}.
\end{eqnarray}
In summary, for $1<\alpha \leqslant 2$ the residence time depends on the measurement time $t$ like
\begin{equation}
 \langle T (0) \rangle \sim \left\{
 \begin{array}{cl}
 t & \mbox{small}\;\;t \\
 \frac{2\Gamma(1/\alpha)L}{\pi(\alpha-1)\sigma_0} \times t^{1-1/\alpha} & \mbox{large}\;\;t \\
 \end{array}
 \right.,
 \label{eq:scaling1a2}
\end{equation}
while for $\alpha<1$
\begin{equation}
 \langle T (0) \rangle \sim \left\{
 \begin{array}{cl}
 t & \mbox{small}\;\;t \\
 \frac{2\Gamma(1-\alpha)}{\pi \alpha} \left(\frac{L}{\sigma_0}\right)^\alpha \sin\frac{\pi\alpha}{2} & \mbox{large}\;\;t \\
 \end{array}
 \right..
 \label{eq:scaling0a1}
 \end{equation}
 For $\alpha=1$, after further calculations, one gets
 \begin{equation}
 \langle T (0) \rangle = t\left[ 1-\frac{2}{\pi}\arctan\frac{\sigma_0 t }{L} \right] + \frac{L}{\pi \sigma_0} \ln\left[ 1+\left(\frac{\sigma_0 t}{L}\right)^2 \right]
 \label{eq:scalinga1}
 \end{equation}
 which for $t\to \infty$ gives logarithmic dependence
 \begin{equation}
 \langle T (0) \rangle \approx \frac{2 L}{\pi \sigma_0} \ln t.
 \end{equation}

\begin{figure}[!ht]
\includegraphics[width=\columnwidth]{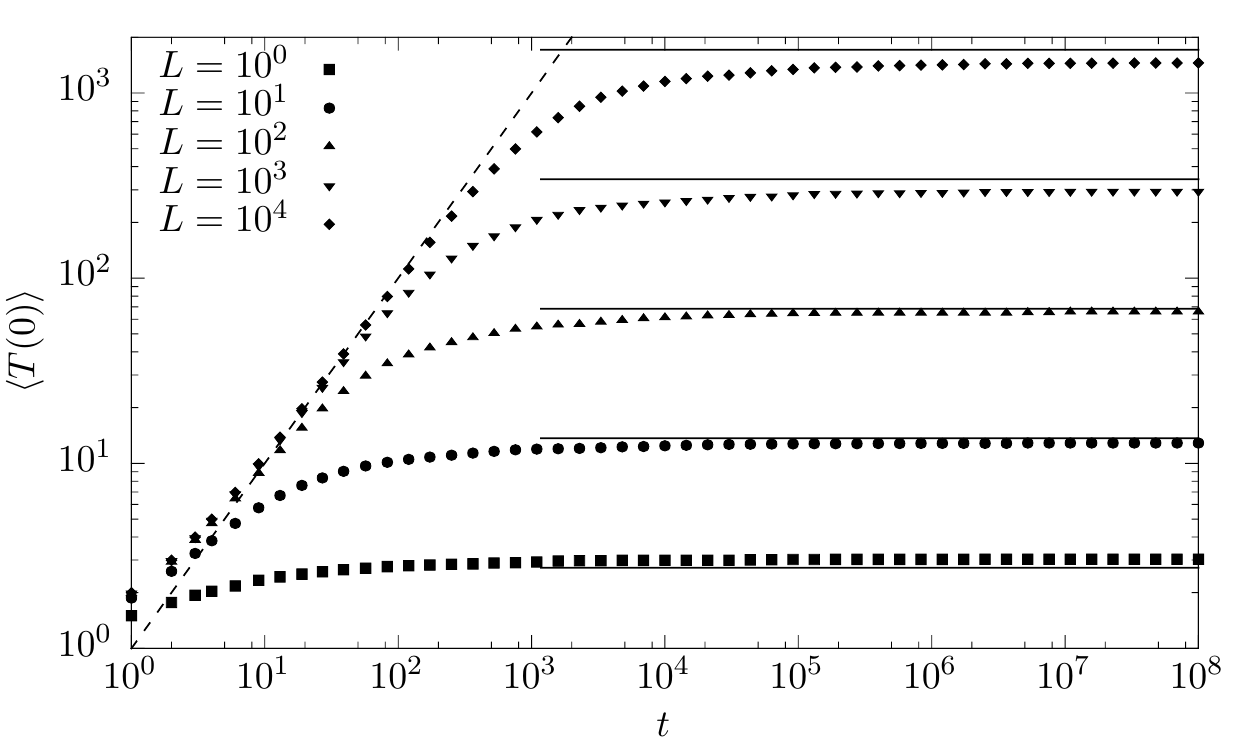} \\
 \includegraphics[width=\columnwidth]{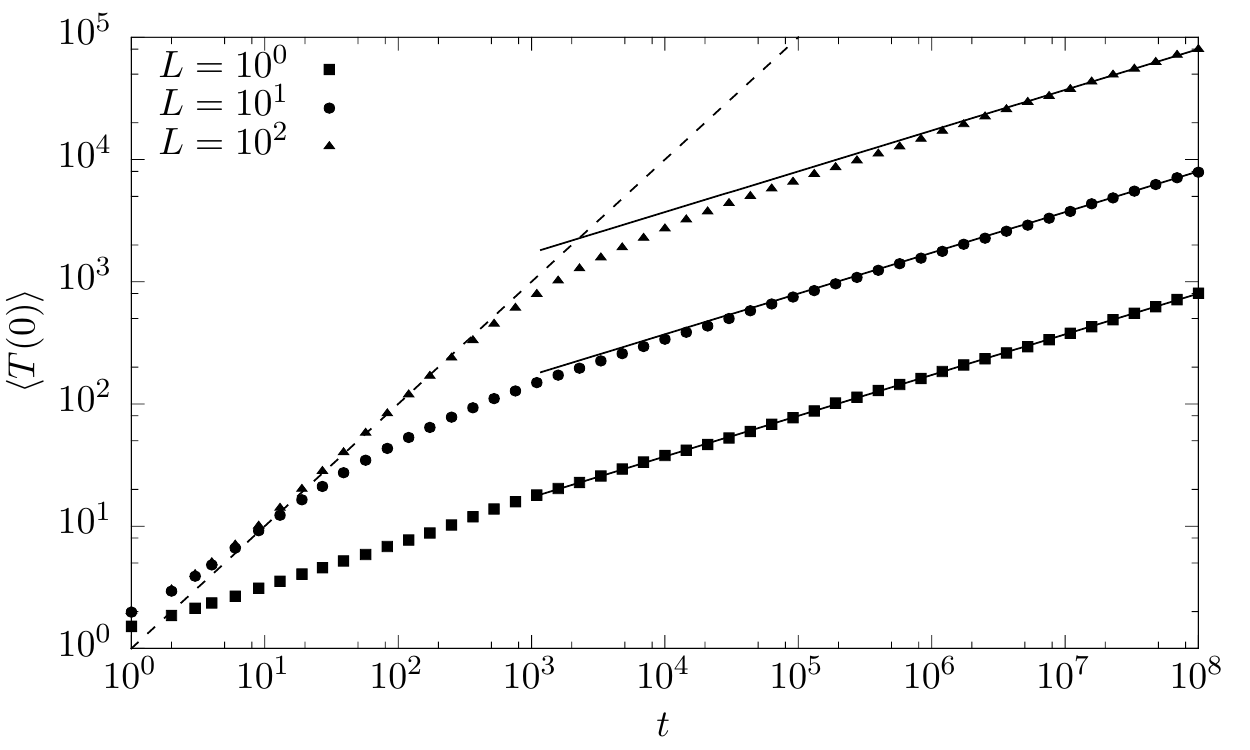} \\
 \includegraphics[width=\columnwidth]{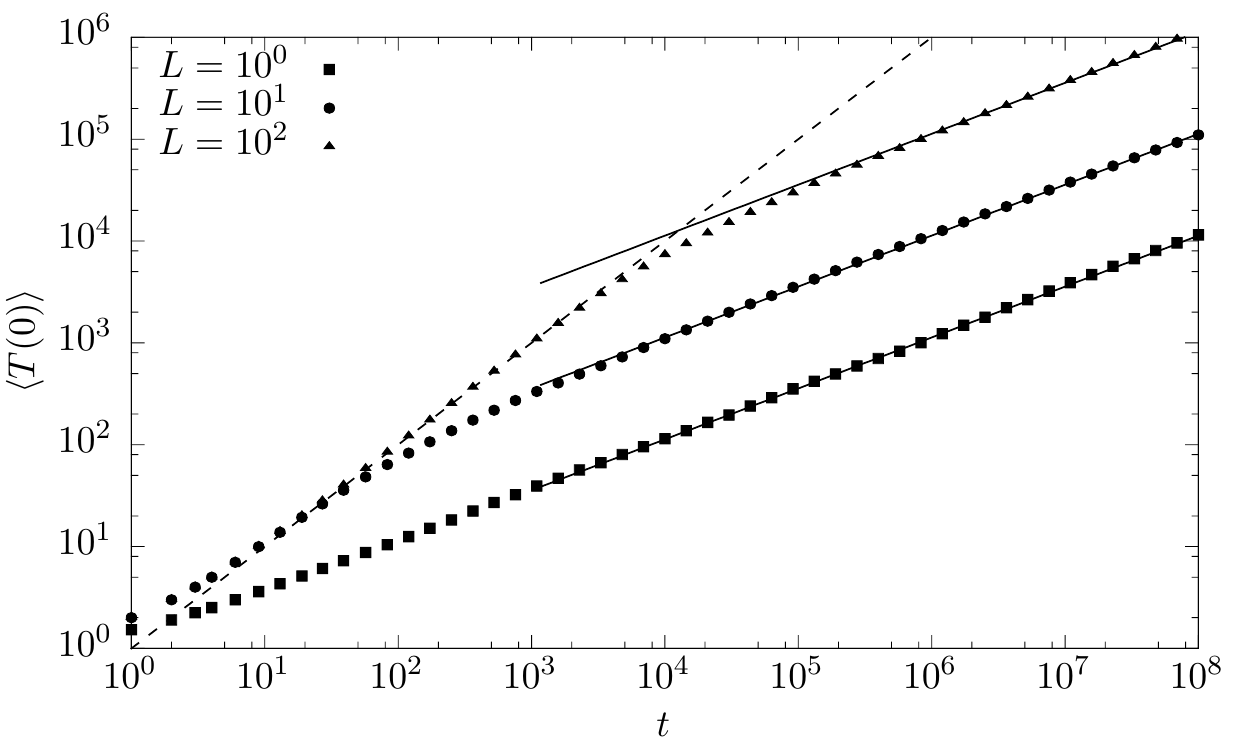}
 \caption{Mean residence time $\langle T(0) \rangle$ for $\alpha=0.7$ (top panel), $\alpha=1.5$ (middle panel) and $\alpha=2$ (bottom panel). Various curves correspond to various interval half-width $L$. Dashed (short time) and solid (long time) lines present scaling given by Eqs.~(\ref{eq:scaling1a2}) and (\ref{eq:scaling0a1}).}
 \label{fig:mrt}
\end{figure}

Figure~\ref{fig:mrt} presents the mean residence time $\langle T(x_0=0) \rangle$ for $\alpha \in \{0.7,1.5,2.0\}$ (from top to bottom) as a function of the measurement time $t$. Various curves correspond to various interval half-width $L$.
The time a particle spends in the interval $[-L,L]$ is always less than the measurement time $t$.
Dashed (short time) and solid (long time) lines in Fig.~\ref{fig:mrt} present scaling of the MRT with the measurement time given by Eq.~(\ref{eq:scaling1a2}) (middle and bottom panels) and Eq.~(\ref{eq:scaling0a1}) (top panel). In both cases simulation results agree well with theoretical predictions. Altogether, we observe that for L\'evy motions with $1<\alpha<2$, the exponent of the asymptotic scaling of MRT is smaller than $1/2$, the value characteristic for a free Brownian diffusion (cf. \cite{barkai2006residence} and Eq.~(\ref{eq:scaling1a2}).
This dependence on the stability index $\alpha$ originates from discontinuity of L\'evy flight trajectories and asymptotics of the first arrival (hitting) time distribution which is of $t^{1/\alpha-2}$ type, cf. \cite{blumenthal1961,chechkin2003b}. For $\alpha<1$ the L\'evy flight process after leaving the $[-L,L]$ interval does not need to return to its interior, see \cite{blumenthal1961}, but it can be jumping back and forth above the $[-L,L]$ interval.
Due to infinite propagation velocity of L\'evy flights such jumps above the interval do not contribute to the residence time.
Consequently, for $\alpha<1$, the mean residence time saturates at the finite value, as in the $\alpha\to 0$ limit discussed above.

\section{L\'evy Walks\label{sec:lw}}

Results presented so far have been constructed for L\'evy flights, when after each time-step $\Delta t$ a random walker performs a jump whose length is distributed according to the $\alpha$-stable density. In this type of the random motion model, the integration time step $\Delta t$ scales the jump length distribution in such a way that for sufficiently small $\Delta t$ resulting mean first passage time and stationary states are invariant with respect to the actual value of $\Delta t$.
On the other hand, propagation of a random walker performing L\'evy flights is characterized by unphysical, infinite velocity. This apparent drawback of the L\'evy flights scenario can be resolved by L\'evy walk models, which in contrast, assume finite velocity of a jump \cite{shlesinger1986,zumofen1993scale}.

Here we use a one-dimensional version of the L\'evy walk model \cite{zumofen1993scale,froemberg2015asymptotic,rebenshtok2014infinite} for which position is a continuous variable:
the jump durations $\mathcal{T}_i$ are set to $\mathcal{T}_i=|\varsigma_i|$ with $\varsigma_i$ being {\it i.i.d} random variables drawn from the symmetric $\alpha$-stable density, see Eq.~(\ref{eq:fcharakt}) and
the jump velocity is characterized by a two-state PDF:
\begin{equation}
 h(v)=\frac{1}{2}\left[ \delta(v-v_0) + \delta(v+v_0) \right],
 \label{eq:jumpvelocity}
\end{equation}
Accordingly,
during each jump a particle position's changes continuously. The jump is finished when the particle travels the total distance $v_0 \mathcal{T}_i$.
After completion of a jump, immediately a new jump duration and a jump velocity are generated.
In order to calculate the position at time $t$, the whole procedure is repeated $n$ times, where $n$ satisfies $\sum_{i=1}^{n-1} \mathcal{T}_i < t < \sum_{i=1}^n \mathcal{T}_i$.
Finally, the position at time $t$ is calculated by adding to the position at time $\sum_{i=1}^{n-1} \mathcal{T}_i$, the velocity $v_0$ multiplied by the time interval $t-\sum_{i=1}^{n-1} \mathcal{T}_i$.
Equivalently, the employed L\'evy walk model can be defined by drawing the jump lengths $\Delta x_i$ from the symmetric $\alpha$-stable density, see Eq.~(\ref{eq:fcharakt}). Negative increments $\Delta x_i$ correspond to jumps taken to the left, while positive $\Delta x_i$ represents jumps to the right. A particle is assumed to move with a constant velocity $v_0$ resulting in the jump duration $\mathcal{T}_i=|\Delta x_i|/v_0$ and
after finishing the jump, a new jump is immediately generated.
Without loss of generality we further set $v_0=1$.
Note, that in the considered L\'evy walk model, the distribution of $\mathcal{T}$ is one-sided, according to the definition $p(\mathcal{T})=p(\varsigma)|d\varsigma/d\mathcal{T}|$ and following Eq.~(\ref{eq:asymptotics}) assumes asymptotically the form $p(\mathcal{T})\approx 2\sigma_0^{\alpha}\sin(\frac{\pi\alpha}{2})\Gamma(\alpha+1)\pi^{-1}\mathcal{T}^{-(1+\alpha)}$.

Unlike for L\'evy flights, the implementation of boundary conditions for L\'evy walks is natural thanks to the continuity of their trajectories. Here, every time a random particle crosses the absorbing boundary, it is removed from the system. At the reflecting boundary, every time a particle hits the point, its trajectory is wrapped, i.e. its motion becomes reversed at the boundary and the random walker continues movement
in the opposite direction.

\subsection{First escape problem\label{sec:feplw}}

\begin{figure}[!ht]
\includegraphics[angle=0, width=\columnwidth]{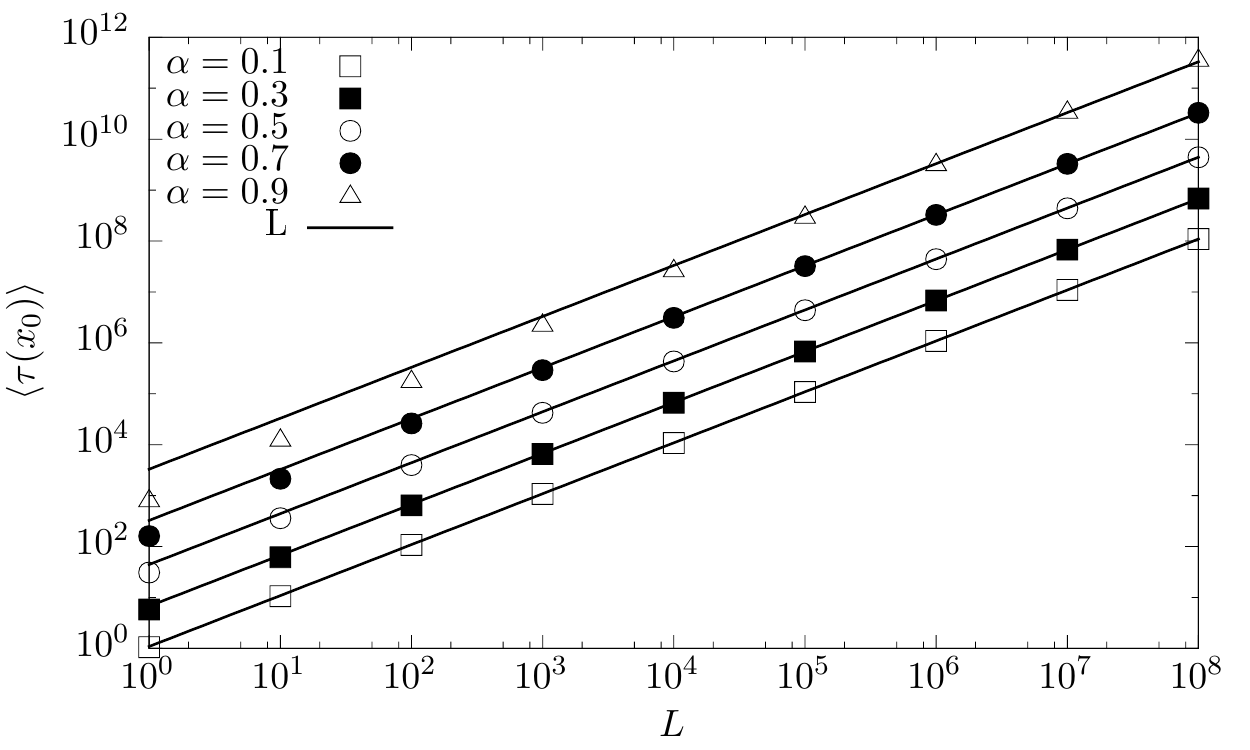}\\
\includegraphics[angle=0, width=\columnwidth]{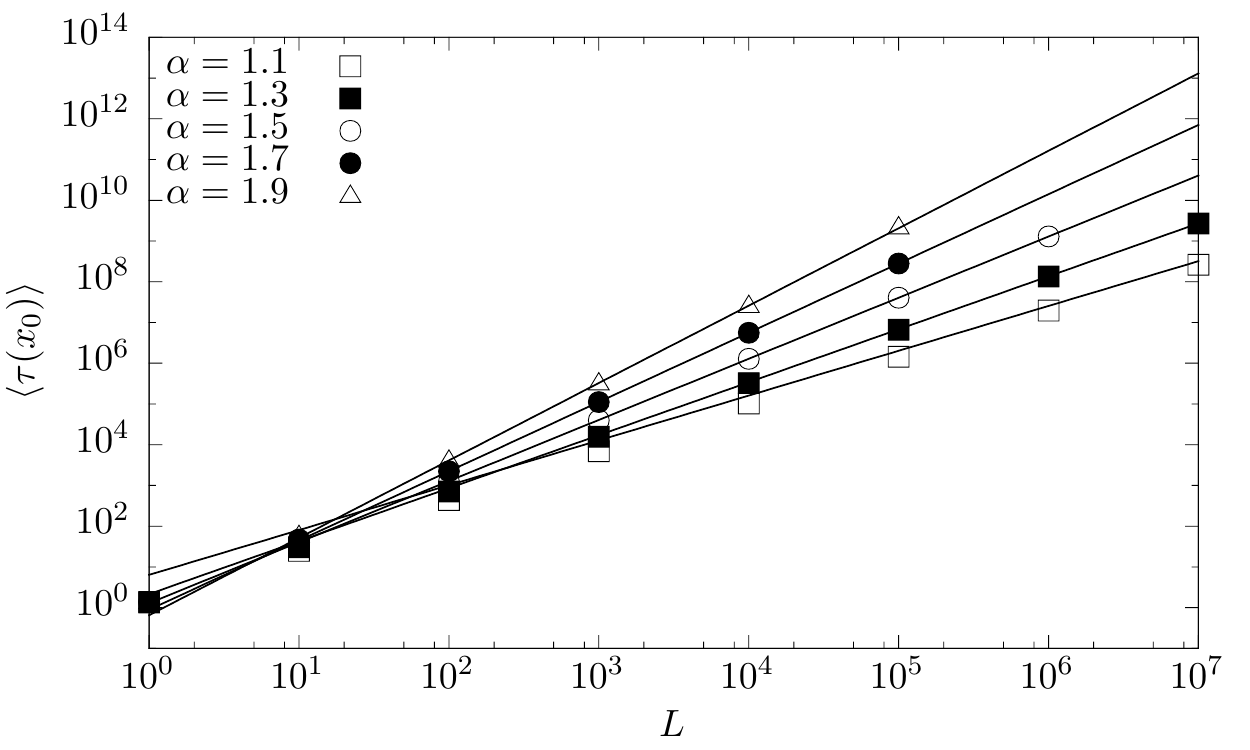}
\caption{Scaling of the mean first passage time on the system half-width $L$ for L\'evy walks restricted by two absorbing boundaries located at $\pm L$ for $\alpha<1$ (top panel) and $\alpha>1$ (bottom panel) with $x_0=0$.
In the top panel, for the clarity of presentation, MFPTs for increasing values of $\alpha$ have been multiplied by a constant factor, otherwise all curves are superimposed.
Solid lines in bottom panel presents $\langle \tau (x_0) \rangle^{\mathrm{LW}}$ calculated according to Eq.~(\ref{eq:kalpha-mfpt}), which perfectly match simulations (points).
}
\label{fig:lw-scaling}
\end{figure}

First we study the problem of the escape from the domain restricted by two absorbing boundaries located at $\pm L$.
In order to verify how the mean first passage time scales with the system size the MFPT has been numerically estimated for a series of increasing interval half-widths $L$.
Fig.~\ref{fig:lw-scaling} shows the MFPT as a function of the interval half width $L$ for $\alpha<1$ (top panel) and $\alpha>1$ (bottom panel). From these plots one can conclude that for large $L$
\begin{equation}
 \langle \tau (0) \rangle \sim
 \left\{
 \begin{array}{ccc}
 L & \mbox{for} & 0<\alpha<1 \\
 L^\alpha & \mbox{for} & 1 < \alpha < 2 \\
 L^2 & \mbox{for} & \alpha \geqslant 2 \\
 \end{array}
 \right..
 \label{eq:mfpt-lw}
\end{equation}
Therefore, for L\'evy walks with $\alpha>1$, the mean first passage time scales on the interval half-width $L$ in the same manner like for L\'evy flights, see Eq.~(\ref{eq:general-mfpt}). Different scaling is recorded for $\alpha<1$, i.e. in the situation when average jump duration/length is infinite, see Eq.~(\ref{eq:mfpt-lw}) and top panel of Fig.~\ref{fig:lw-scaling}.

\begin{figure}[!ht]
\includegraphics[angle=0, width=\columnwidth]{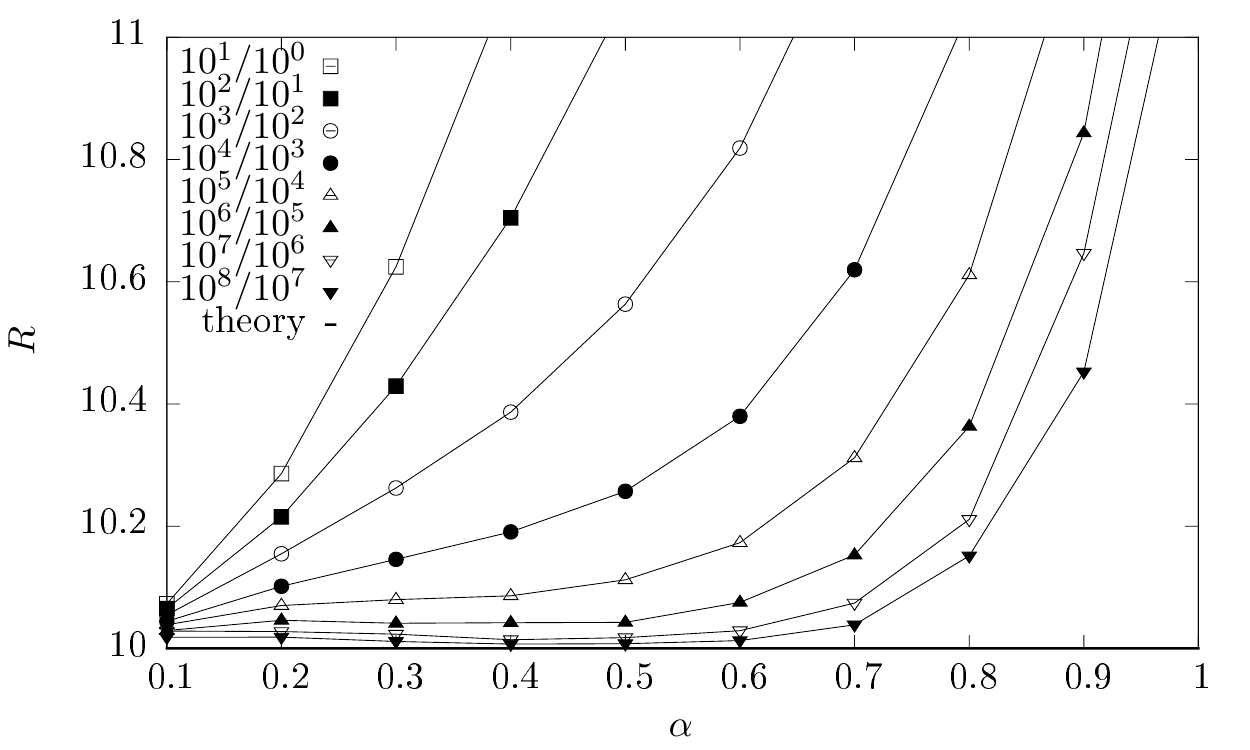}\\
\includegraphics[angle=0, width=\columnwidth]{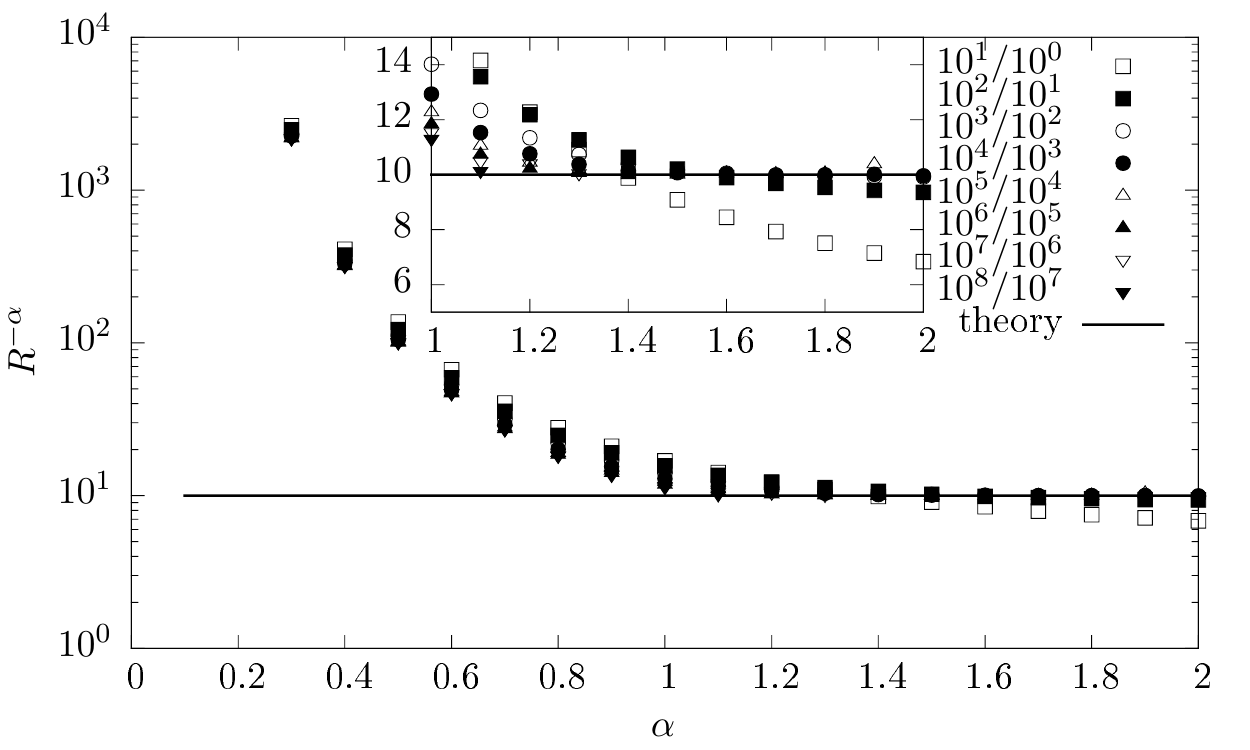}
\caption{Ratio $R$, see Eq.~(\ref{eq:ratio}), of the mean first passage times for L\'evy walks restricted by two absorbing boundaries located at $\pm L$.
Asymptotically, for $\alpha<1$ one has $R=10$, while for $\alpha>1$ one gets $R=10^\alpha$. In the bottom panel, the inset enlarges $1<\alpha<2$ region.
}
\label{fig:lw-ratio}
\end{figure}

In order to improve the clarity of the presented figure, in the top panel of Fig.~\ref{fig:lw-scaling}, curves corresponding to increasing values of $\alpha$ have been shifted upwards by multiplying MFPTs by a constant factor, otherwise they are superimposed.
Furthermore, from obtained $\langle \tau (x_0) \rangle$ the ratio of mean first passage times $\langle \tau (x_0) \rangle$s has been calculated
\begin{equation}
 R=\frac{\langle \tau (x_0) \rangle_{10L}}{\langle \tau (x_0) \rangle_{L}}.
 \label{eq:ratio}
\end{equation}
Fig.~\ref{fig:lw-ratio} presents $R$ for $\alpha<1$ and $R^{-\alpha}$ for $\alpha>1$ (with $x_0=0$) which in more details demonstrate how scaling from Eq.~(\ref{eq:mfpt-lw}) is reached.
From Eq.~(\ref{eq:mfpt-lw}) it implies that $R \approx 10$ (for $\alpha<1$) and $R^{-\alpha} \approx 10$ (for $\alpha>1$).
Indeed such a behavior is visible in Fig.~\ref{fig:lw-ratio} especially for large $L$.

Equation~(\ref{eq:mfpt-lw}) and Fig.~\ref{fig:lw-scaling} suggest that the mean first passage time for L\'evy walks with $1 < \alpha \leqslant 2$ scales similarly like for L\'evy flights, i.e. $\langle \tau (0) \rangle \sim L^\alpha$, see Eq.~(\ref{eq:general-mfpt}).
As mentioned above, a main difference between these two categories of free motion is in the finite propagation velocity $v_0$ and continuity of trajectories for LW, versus infinite propagation velocity and discontinuity of trajectories for LF. In both cases however, jumps are distributed with an $\alpha$-stable density. Therefore, one can expect that for large interval half-width $L$, L\'evy walks can be effectively approximated by L\'evy flights with some unknown scale parameter $\sigma_0^{\mathrm{LW}}$ or anomalous diffusion constant $K_\alpha^{\mathrm{LW}}=\left[ \sigma_0^{\mathrm{LW}} \right]^\alpha$, see Eq.~(\ref{eq:ffpe}). At the same time the finite propagation introduces cut-off for time dependent densities for LW model. Their support is restricted to $[-v_0 t, v_0 t]$.

The anomalous diffusion constant $K_\alpha^{\mathrm{LW}}$ can be estimated by \cite[Eqs.~(4) and (42)]{rebenshtok2014infinite} and Eqs.~(\ref{eq:asymptotics}) and~(\ref{eq:jumpvelocity}) resulting in
\begin{equation}
 K_\alpha^{\mathrm{LW}} = \frac{ 2 \sigma_0^\alpha \Gamma(1+\alpha) |\Gamma(-\alpha)| \sin \frac{\pi \alpha}{2} \left| \cos \frac{\pi \alpha}{2} \right| |v_0|^\alpha}{ \langle \mathcal{T} \rangle \pi},
 \label{eq:kth}
\end{equation}
where $\langle \mathcal{T} \rangle$ is the mean waiting time for a next jump (flight duration) and $\sigma_0$ is the scale parameter in Eq.~(\ref{eq:fcharakt}) and $v_0=1$. $\langle \mathcal{T} \rangle$ is finite for $1 < \alpha \leqslant 2$.
Application of Eq.~(\ref{eq:kth}) requires knowledge of $\langle \mathcal{T} \rangle$ which can be estimated numerically or with the help of \cite[Eq.~(77)]{rebenshtok2014infinite}
\begin{equation}
 \langle \mathcal{T} \rangle = 2\sigma_0 \frac{\Gamma(1-1/\alpha)}{\pi}.
 \label{eq:tau}
\end{equation}
Formula~(\ref{eq:tau}) perfectly approximates $\langle \mathcal{T} \rangle$.
For $\alpha \geqslant 1.2$ errors are smaller than $0.3\%$ of the exact value.
For $\alpha \to 1$, approximation breaks down reflecting the fact that $\langle \mathcal{T} \rangle$ diverges.
Finally one gets the formula for $K_\alpha^{\mathrm{LW}}$
\begin{equation}
 K_\alpha^{\mathrm{LW}} = \frac{ \sigma_0^{\alpha-1} \Gamma(1+\alpha) |\Gamma(-\alpha)| \sin \frac{\pi \alpha}{2} \left| \cos \frac{\pi \alpha}{2} \right| |v_0|^\alpha}{ \Gamma(1-1/\alpha) }.
 \label{eq:kalphatheory}
\end{equation}

If the L\'evy flight approximation to L\'evy walks works, the mean first passage time for LW could be estimated from the formula analogous to Eq.~(\ref{eq:general-mfpt})
\begin{equation}
 \langle \tau (0) \rangle^{\mathrm{LW}}=
 \frac{L^\alpha}{K_\alpha^{\mathrm{LW}} \Gamma(1+\alpha)}.
 \label{eq:kalpha-mfpt}
\end{equation}
Estimated values of $K_\alpha^{\mathrm{LW}}$, see Eq.~(\ref{eq:kalphatheory}), are given in Tab.~\ref{tab:sigma}.
These values have been used to calculate the MFPT according to Eq.~(\ref{eq:kalpha-mfpt}), see solid lines in the bottom panel of Fig.~\ref{fig:lw-scaling}.
Therefore, solid lines in the bottom panel of Fig.~\ref{fig:lw-scaling} not only display the $L^\alpha$ scaling, but also the MFPT values estimated by the LF approximation to LW, see Eq.~(\ref{eq:kalpha-mfpt}). The latter works nicely for large $L$ and $\alpha>1$.

\begin{table}[!h]
 \begin{tabular}{l||c|c|c|c|c}
$\alpha$ & 1.1 & 1.3 & 1.5 & 1.7 & 1.9 \\ \hline
$K_\alpha^{\mathrm{LW}}\;\;\;$ & 0.1495 & 0.3981 & 0.5863 & 0.7294 & 0.8401 \\
\end{tabular}
\caption{Values of $K_\alpha^{\mathrm{LW}}$ calculated according to Eq.~(\ref{eq:kalphatheory}).}
\label{tab:sigma}
\end{table}

\begin{figure}[!ht]
\includegraphics[angle=0, width=\columnwidth]{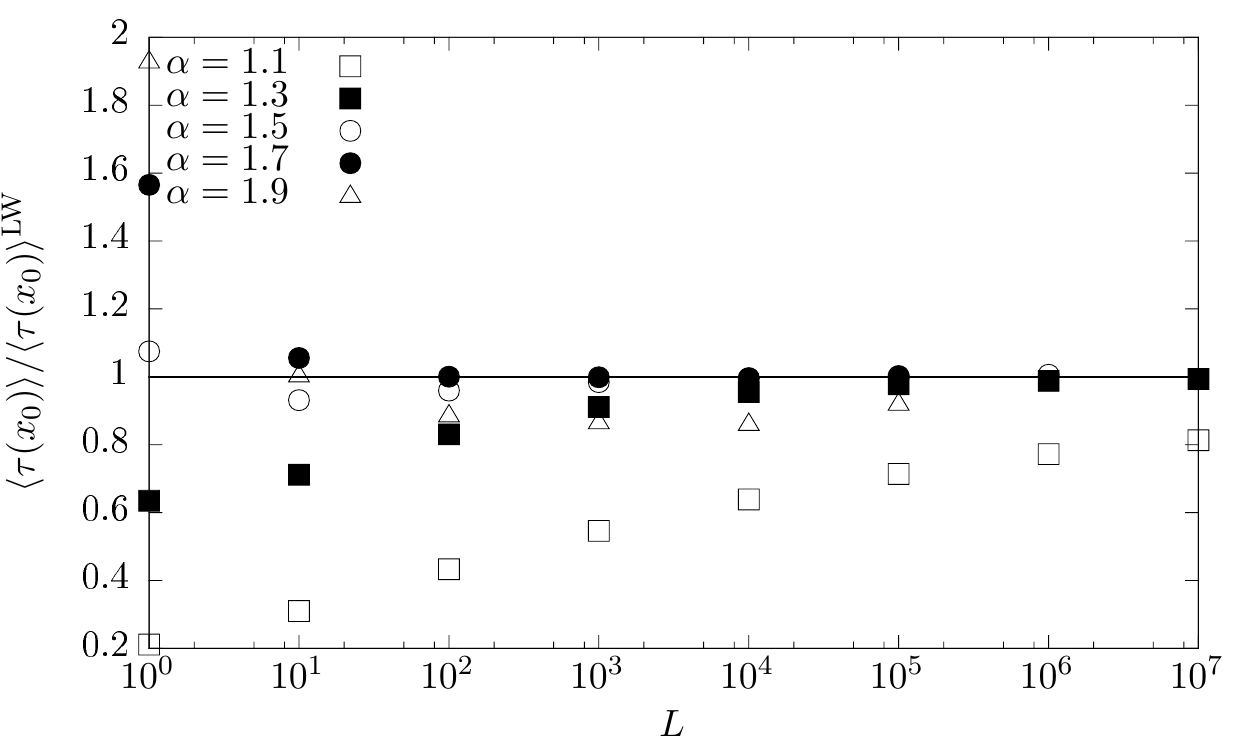}\\
\caption{Ratio $\tilde{R}=\langle \tau (x_0) \rangle / \langle \tau (x_0) \rangle^{\mathrm{LW}} $ of numerically estimated mean first passage time for LW for $x_0=0$ and the mean first passage time calculated according to Eq.~(\ref{eq:kalpha-mfpt}).
}
\label{fig:mfpt-sigma}
\end{figure}

In order to measure the quality of LF approximation to LW, the ratio of numerically estimated MFPT for LW,
$\langle \tau (0) \rangle$, and value of the MFPT evaluated according to
Eq.~(\ref{eq:kalpha-mfpt}),
$\langle \tau (0) \rangle^{\mathrm{LW}}$, has been introduced:
\begin{equation}
\tilde{R}=\langle \tau (0) \rangle / \langle \tau (0) \rangle^{\mathrm{LW}}
\end{equation}
and analyzed for various ranges of $L$ and $\alpha$, see Fig.~\ref{fig:mfpt-sigma}.
With the increasing interval half width $L$ the quality of LF approximation improves resulting in $\tilde{R}\approx 1$. Moreover, approximation of L\'evy walks by L\'evy flights works better for larger values of the stability index $\alpha$.

Finite propagation velocity results in finite support for LW which is restricted to the interval $[-v_0 t,v_0 t]$. At the same time L\'evy flights are unconstrained and located at any point on the real line as they propagate with the infinite velocity. Despite this fundamental difference in the propagation velocity
L\'evy flights seem to well approximate L\'evy walks in the central part of the respective propagator (time-dependent PDF).
For free L\'evy flights the scale parameter, which defines central part of the distribution, grows like $\sigma_0^{\mathrm{LF}} t^{1/\alpha}$, where $\sigma_0^{\mathrm{LF}}$ is the scale parameter of the jump length density, see Eq.~(\ref{eq:sigma}). The same dependence of the scale parameter is observed for free L\'evy walks, i.e. $\sigma_0^{\mathrm{LW}} t^{1/\alpha}$.

\begin{figure}[!ht]
\includegraphics[angle=0, width=\columnwidth]{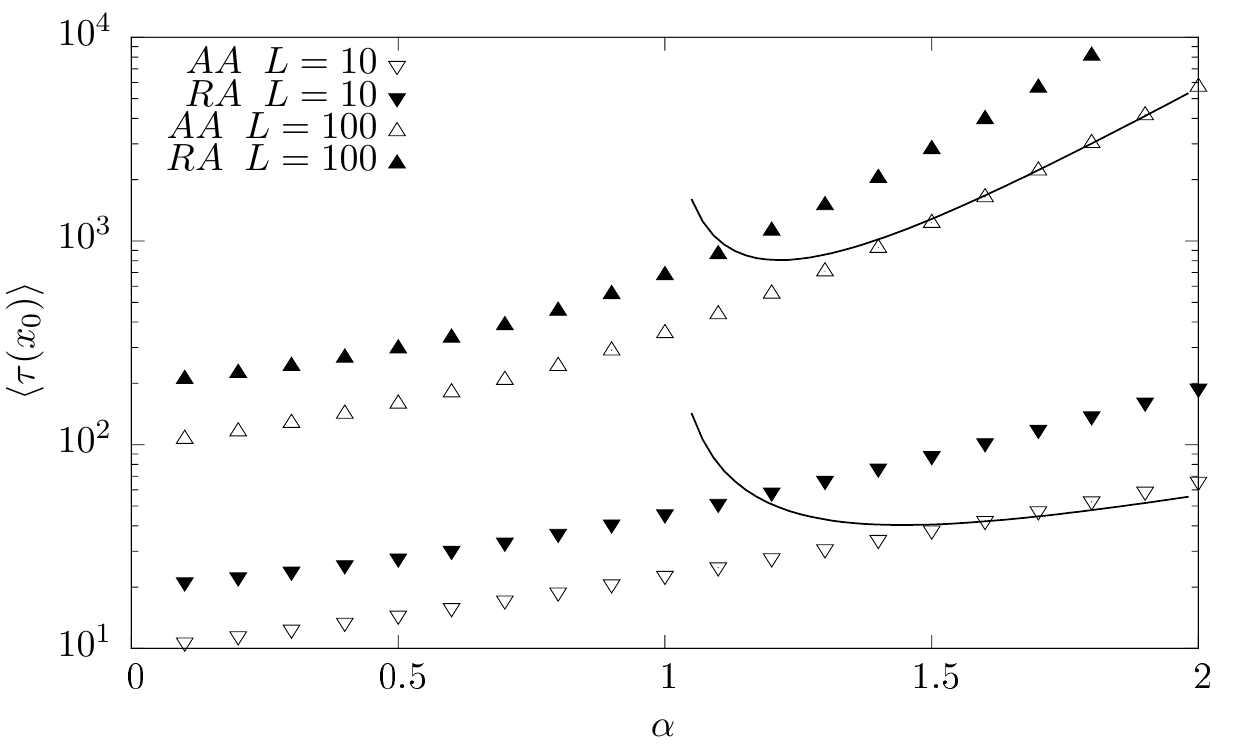}\\
\includegraphics[angle=0, width=\columnwidth]{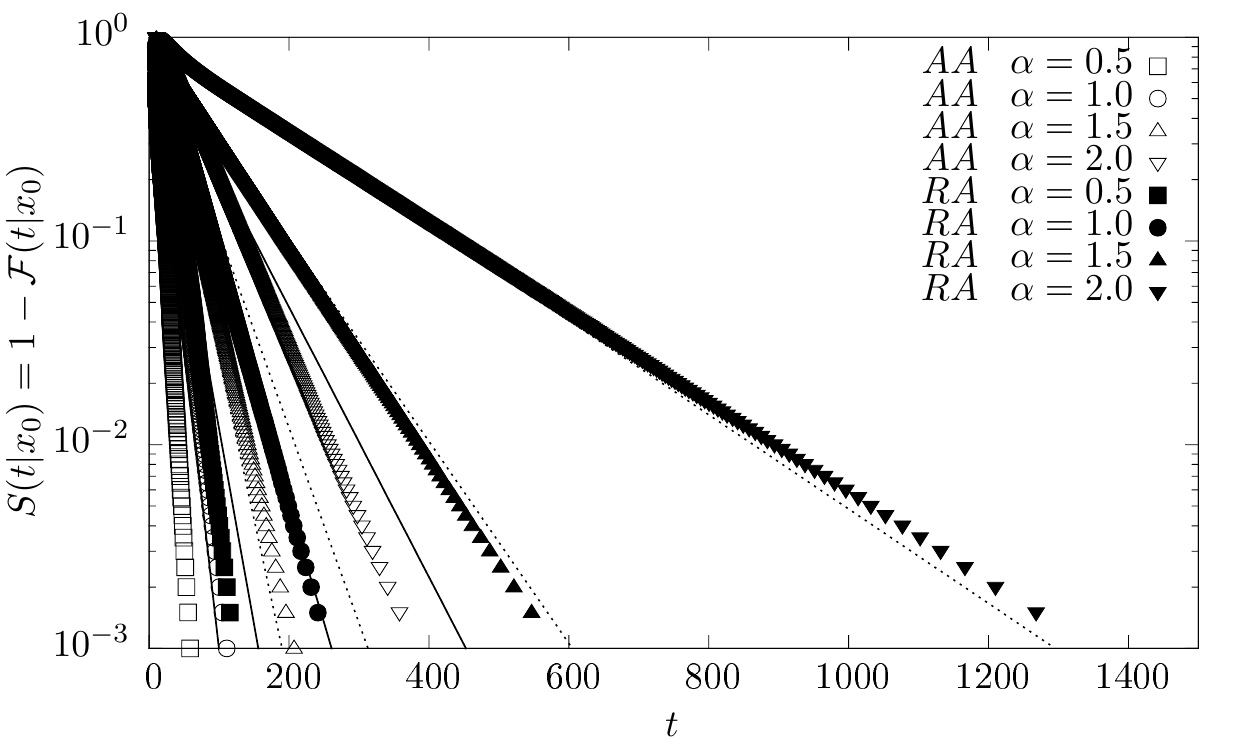}\\
\includegraphics[angle=0, width=\columnwidth]{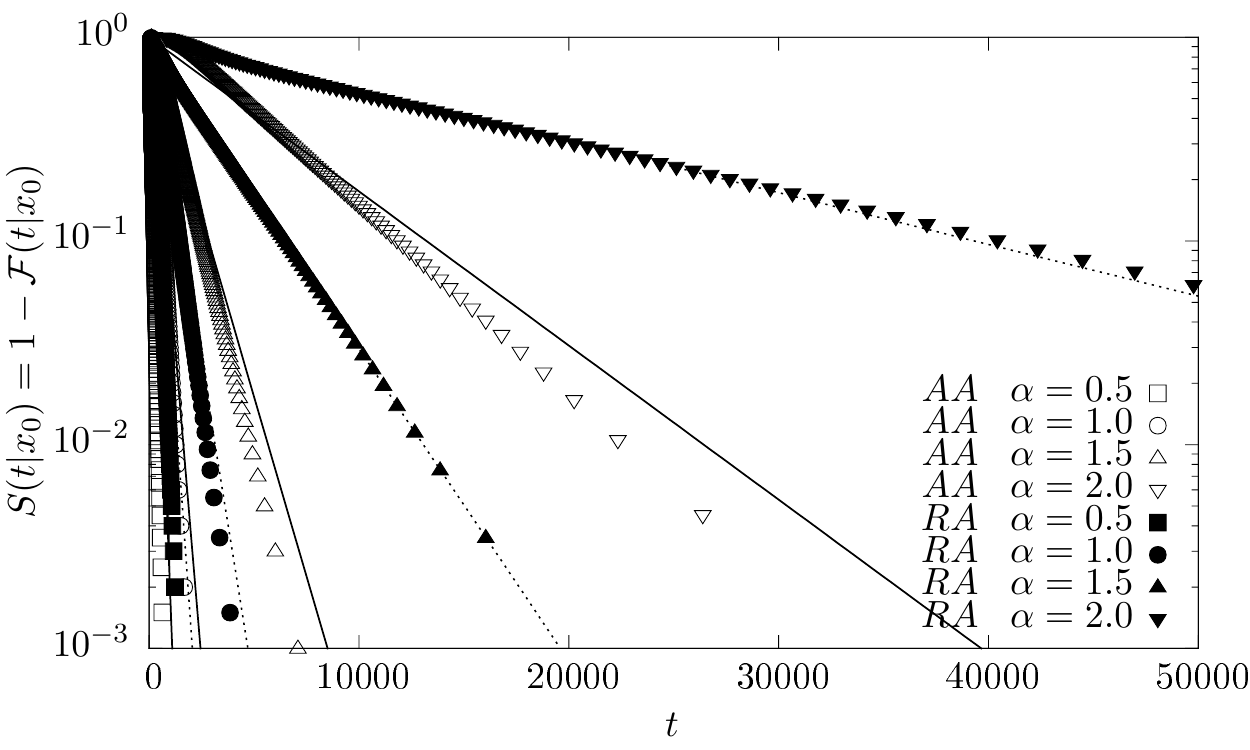}\\
\caption{Mean first passage time $\langle \tau (x_0)\rangle $ for L\'evy walks from the interval restricted by two absorbing-absorbing or reflecting-absorbing boundaries located at $\pm L$ (top panel).
Solid lines present approximation given by Eq.~(\ref{eq:kalpha-mfpt}).
Middle and bottom panels present survival probabilities for absorbing-absorbing (empty symbols) and reflecting-absorbing (full symbols) boundary conditions with $L=10$ (middle panel) and $L=100$ (bottom panel).
\textcolor{black}{
Lines present $S(t|x_0) \approx \exp\left[ -t/\langle \tau (x_0) \rangle \right]$ approximation for absorbing-absorbing (solid lines) and reflecting-absorbing (dotted lines) boundary conditions.
}
Results are averaged over $N=10^7$ realizations with $x_0=0$.
}
\label{fig:lw-l10}
\end{figure}

Figure~\ref{fig:lw-l10} presents MFPT for L\'evy walks with $L=10$ and $L=100$ as a function of the stability index $\alpha$.
Middle and bottom panels present survival probabilities $S(t|x_0)$ with $x_0=0$ for $L=10$ and $L=100$, respectively.
From the top panel it is clearly visible that approximation~(\ref{eq:kalpha-mfpt}) works better for large interval half-width $L$. This is in line with previous considerations.
Analogously like for L\'evy flights, for a fixed interval half width $L$, the mean first passage time for reflecting-absorbing scheme is larger than for absorbing-absorbing scenario because in the former scenario distance to an absorbing boundary is larger than in the latter scheme.
Due to finite particle velocity, for a fixed $L$, the mean first passage time for reflecting-absorbing scenario is not equal to the MFPT for absorbing-absorbing scenario with two times wider interval half-width (results not shown) as it was observed for L\'evy flights in the wrapping scenario.
Like for L\'evy flights, see Fig.~\ref{fig:aa-fpt}, survival probabilities have exponential tails. Empty symbols in middle and bottom panels of Fig.~\ref{fig:lw-l10} correspond to the absorbing-absorbing scenario, while full symbols to the reflecting-absorbing scenario. 
\textcolor{black}{
At the same time lines present $S(t|x_0)\approx \exp\left[ -t/\langle \tau (x_0) \rangle \right]$ approximation to the survival probability for absorbing-absorbing (solid lines) and reflecting-absorbing (dotted lines) boundary conditions. Quality of the approximation depends in non trivial way on the stability index $\alpha$ and type of boundary conditions. In particular, such an approximation works better for reflecting-absorbing than absorbing-absorbing setup.
The discrepancy between the actual decay and $\exp\left[ - t /\langle \tau (x_0) \rangle \right]$ means that the exponential decay is only asymptotic, since if we had an exponential decay for all times, the law (\ref{eq:approximation}) should hold.
}
Finally, the distribution of number of jumps performed until leaving of the $[-L,L]$ interval has exponential asymptotics (results not shown).


\subsection{Stationary states\label{sec:stationarylw}}

\begin{figure}[!ht]
\includegraphics[angle=0, width=\columnwidth]{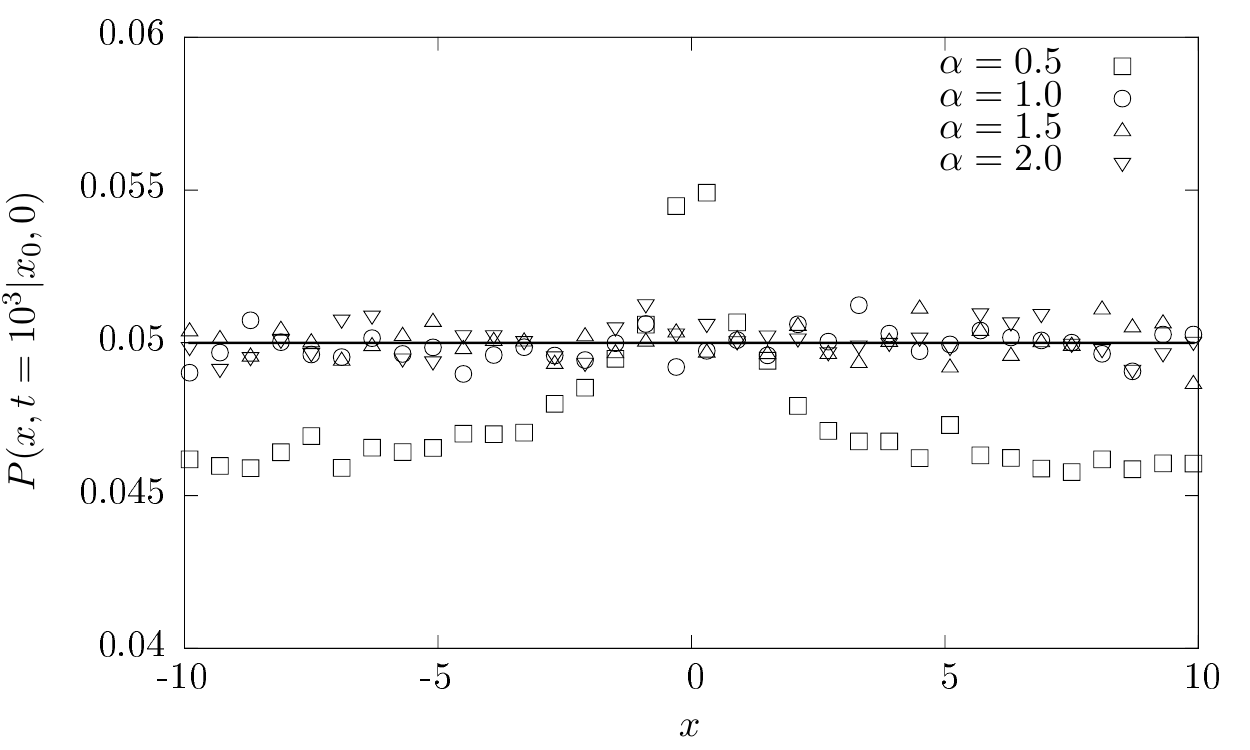}\\
\includegraphics[angle=0, width=\columnwidth]{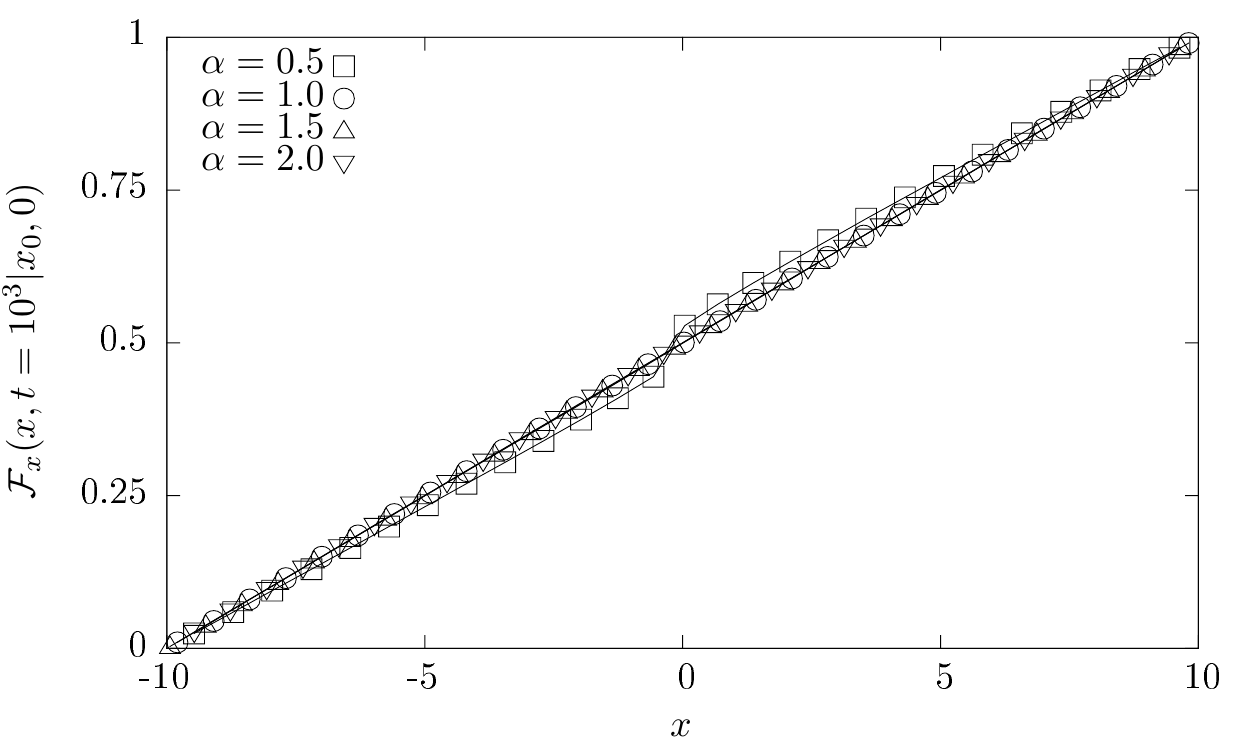}
\caption{Probability densities $P(x,t=10^3|x_0,0)$ (top panel) and corresponding cumulative densities $\mathcal{F}_x(x)$ (bottom panel) for L\'evy walks restricted by two reflecting boundaries located at $\pm 10$.
Results are averaged over $N=10^7$ realizations with $x_0=0$.
}
\label{fig:lw-stationary}
\end{figure}

Figure~\ref{fig:lw-stationary} presents histograms for L\'evy walk confined by two reflecting boundaries located at $\pm 10$ (top panel) and corresponding cumulative densities $\mathcal{F}_x(x)$ (bottom panel) at $t=10^3$. For $\alpha>1$, the system already reached its stationary state which is a uniform probability density $P_{st}(x)=\frac{1}{2L}$, while for $\alpha<1$ a persistent cusp at the $x_0=0$ (origin) is still visible. With increasing time $t$ the hight of this peak is decreasing and time dependent density become more and more uniform.
In the probability density, the peak visible for $\alpha=0.5$ is responsible for the jump of the cumulative density at $x\approx 0$.

\begin{figure}[!ht]
\begin{tabular}{cc}
\includegraphics[angle=0, width=0.5\columnwidth]{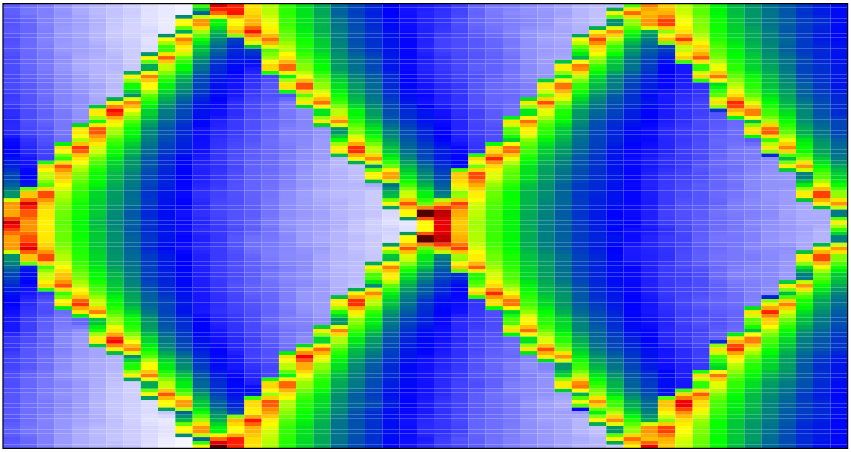} &
\includegraphics[angle=0, width=0.5\columnwidth]{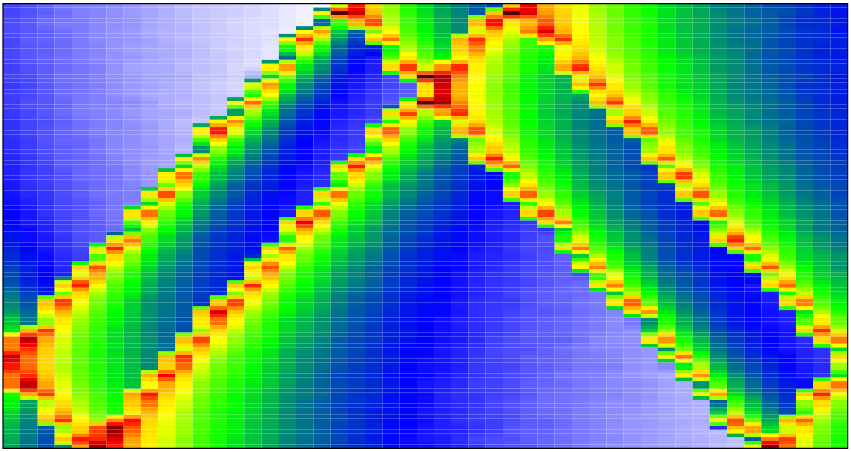} \\
\end{tabular}
\includegraphics[angle=0, width=\columnwidth]{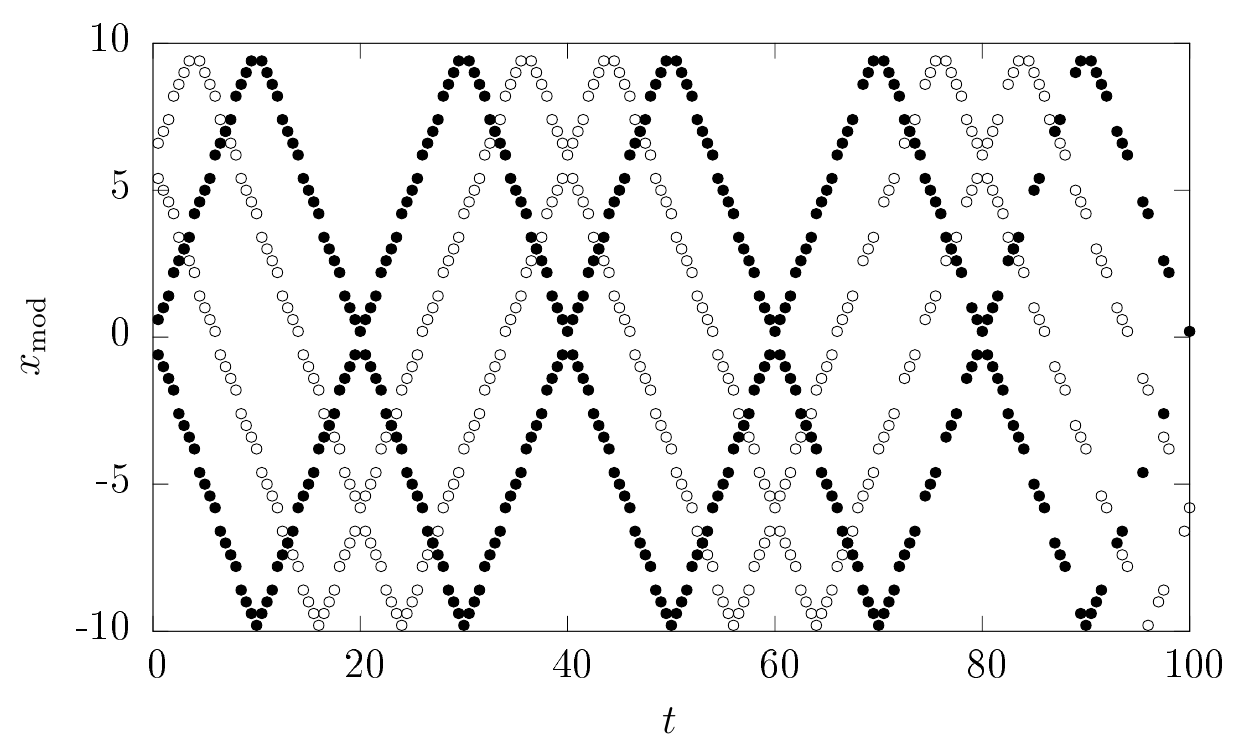}\\
\includegraphics[angle=0, width=\columnwidth]{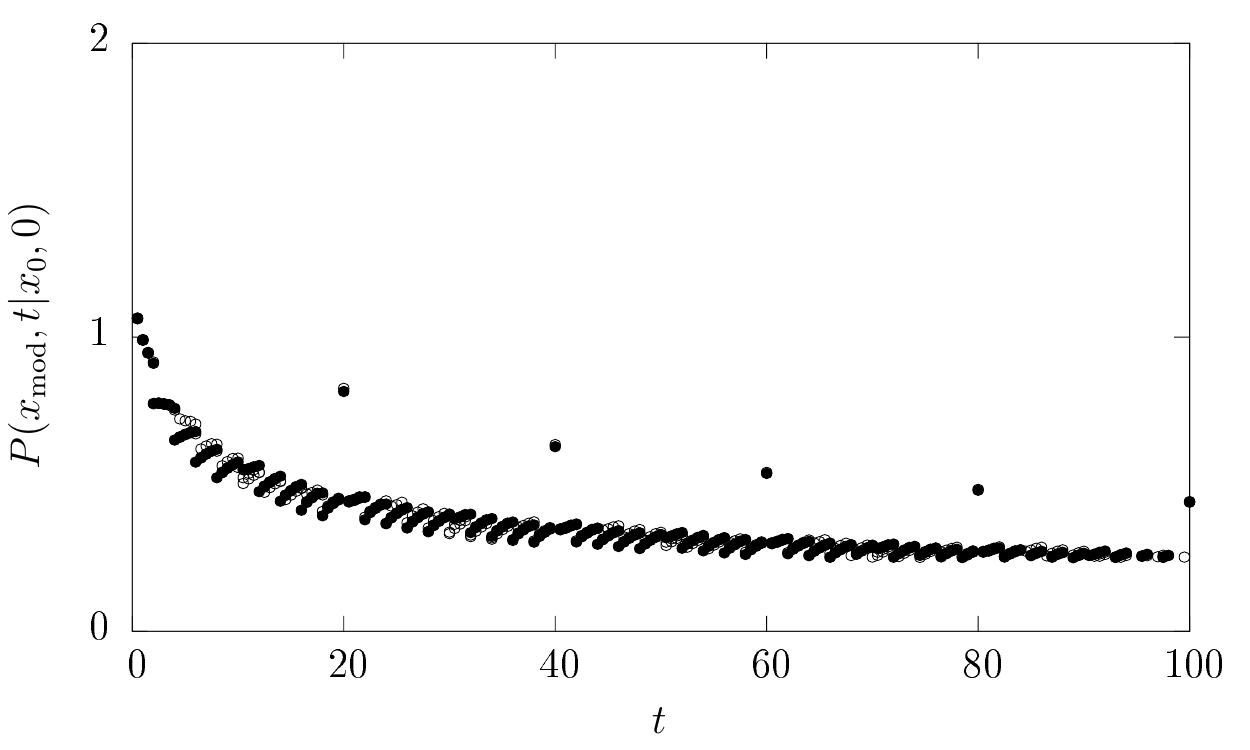}\\
\caption{
Sample heat-maps presenting time dependent densities $P(x,t|x_0,0)$ for $\alpha=0.5$ with $60 \leqslant t \leqslant 100$ for $x_0=0$ (top left) and $x_0=6$ (top right).
Middle panel presents position of $P(x,t|x_0,0)$ maxima for $\alpha=0.5$ with $x_0=0$ (filled dots) and $x_0=6$ (empty dots). The bottom panel shows maximal values of $P(x,t|x_0,0)$.
Multiple points for the same $t$ correspond to multi-modal time dependent densities.
}
\label{fig:lw-timedependent}
\end{figure}

Figure~\ref{fig:lw-timedependent} present sample time dependent densities $P(x,t|x_0,0)$ as heat-maps (top panel), location of maxima of $P(x,t|x_0,0)$ (middle panel) and maximal values of $P(x,t)$ (bottom panel).
Maxima of probability densities originate due to the initial condition, i.e. $x(0)=x_0$. Their height is a decaying function of time with the decay rate dependent on the stability index $\alpha$.
The slowest decay is observed for $\alpha<1$, when the average jump length is infinite, because the initial peak has smallest number of chances to bifurcate.
Putting it differently, height of the maximum of probability density decreases every time a jump direction and jump length are generated.
Therefore, the number of jumps (determined by $\alpha$) defines the decay rate of the initial condition.
Fig.~\ref{fig:lw-timedependent} presents results for $\alpha=0.5$ with $x_0=0$ and $x_0=6$ because such a choice of parameters nicely shows general properties of time dependent densities $P(x,t|x_0,0)$.
Analogously, positions of maxima are not fixed but they constantly, ballistically move and bifurcate due to finite propagation velocity $v_0=1$ and equal probability of jumps to the left and right.
The initial cusp splits into two parts moving to the right and left.
For $\alpha<1$ average jump duration is infinite, therefore these parts return to the initial position after $\Delta t=2L/v_0$ (for $x_0=0$) or $\Delta t=4L/v_0$ (for $x_0 \neq 0$). This effect is nicely visible in the middle panel of Fig.~\ref{fig:lw-timedependent}.
Multiple points for the same $t$ in the bottom panel correspond to bimodal time dependent densities.
The non-monotonous, point-like amplification of peaks takes place when two peaks moving in opposite directions meet in one point. At this point height of the peak is twice the background.
In the limit of large times $t$ time dependent densities converge to the uniform stationary density, i.e. $P_{st}(x)=1/2L$.

\section{Summary and conclusions \label{sec:summary}}

Both L\'evy walks and L\'evy flights assume that a random walker performs long jumps distributed according to a heavy-tailed, power-law density. The main difference between both scenarios is in continuous trajectories and finite propagation velocity of L\'evy walks versus discontinuous trajectories and infinite propagation velocity of L\'evy flights. Nevertheless, due to the same type of the jump length density both scenarios are deeply related and need to be compared in great detail.
Such a comparison between behavior of LF and LW, with respect to a class of important observables like mean first passage time, survival probabilities and stationary states provided the main motive of current research. We have verified when both models are similar and when they differ.
In addition to comparison of two classical random walk schemes we studied the problem of posing boundary conditions for L\'evy flights and L\'evy walks.

The mean first passage time for L\'evy flights scales asymptotically as $L^\alpha$ with the interval-half width, what is especially well visible for $x_0=0$.
The same scaling is observed for L\'evy walks with $\alpha>1$ and large interval half-width $L$. For $\alpha<1$ the mean first passage time is proportional to $L$.
\bd{Eq.~(\ref{eq:mfpt-lw}), which is the main outcome of this part of the manuscript,}
 indicates that for $\alpha>1$ L\'evy flights with properly adjusted scale parameter can approximate L\'evy walks with respect to the analysis of mean first passage time.

In addition to the escape of L\'evy flights from finite intervals we have analyzed the problem of the residence time, i.e. the fraction of time which unbounded process spends in the prescribed part of the line, e.g. a finite interval.
The mean residence time of LF as a function of the measurement time $t$ has a universal short time scaling $\langle T (0) \rangle \sim t$ and a non-universal long time scaling $t^{1-1/\alpha}$ (for $1 < \alpha \leqslant 2$). For $\alpha<1$, due to discontinuity of trajectories of L\'evy flights and their transient character, the mean residence time saturates at a constant value as measurement time goes to infinity.
\bd{Results regarding the mean residence time, Eq.~(\ref{G-13}), and dependence of the mean residence time on the measurement time $t$, Eqs.~(\ref{eq:scaling1a2})--(\ref{eq:scalinga1}), constitute key results of this part of the manuscript.}

Problem of formulation of boundary conditions on a single trajectory level for L\'evy flights is not fully resolved. While it is known how to treat the absorbing boundaries, it is not uniquely defined how to implement reflecting boundary conditions. We have compared two scenarios: wrapping (reversing) and stopping (pausing) showing that the former results in uniform stationary states for a motion restricted by two reflecting boundaries, while the latter scenario gives the same stationary states as impenetrable boundary conditions \cite{denisov2008} and steep potential wells \cite{dubkov2007}. In the case of L\'evy walks, the problem of posing reflecting boundary conditions is more apparent. Finite propagation velocity and continuous paths suggest that trajectory should be wrapped along reflecting boundaries. As a consequence stationary states for a motion restricted by two reflecting boundaries are uniform. Nevertheless, L\'evy walks display slowly decaying memory about initial conditions especially in situations when mean jump length (flight time) is divergent.

\bd{
Our work clarifies the cases where L\'evy flights can be used as an approximation to L\'evy walks.
Finite propagation velocity of LW makes them inertial, i.e. a test particle moves with finite kinetic energy, and collisions leading to instantaneous alternations  of velocities are fully elastic. In consequence LW trivially fulfills Newton equations. This is not however the case of LFs: 
Inertial L\'evy flights \cite{sokolov2010,lu2011inertial} break the equipartition theorem and lead to inter-dependence between the position and velocity, which is manifested by the nontrivial joint distribution of both variables. Moreover, in view of thermodynamic interpretation, the Langevin equation with non-Gaussian L\'evy noise, Eq.~(\ref{eq:langevin2}) describes change of the position of a particle under the action of nonequilibrated  external forcing which does not fulfill the standard fluctuation-dissipation theorem \cite{dybiec2012,kusmierz2014}.}

\bd{Within the effort of explaining real life phenomena exhibiting anomalous diffusion, Langevin or generalized Langevin description is frequently a natural choice \cite{zaburdaev2015levy,Lubashevsky2009,Ferreira2012,Kneller2011,Bao2006}. However, a full correspondence of that picture with L\'evy walks and L\'evy flights dynamics still calls for further systematic studies.}

\section{Acknowledgments}

This project has been supported in part (BD and EGN) by the grant from National Science Center (2014/13/B/ST2/02014).
EB thanks the Israel Science Foundation for support.
Computer simulations have been performed at the Academic
Computer Center Cyfronet, Akademia G\'orniczo-Hutnicza (Krak\'ow, Poland).

%
%

\def\url#1{}\def\url#1{}

\end{document}